\documentclass[aps,prd,showpacs,twocolumn,superscriptaddress,nofootinbib,preprintnumbers]{revtex4-1} 

\makeatletter
\def\p@subsection{}
\makeatother

\usepackage{color,graphicx,amsmath,amssymb,lipsum,bm}
\usepackage[colorlinks=true,citecolor=blue,linkcolor=blue,urlcolor=blue, backref=false,pdfborder={0 0 0}]{hyperref}

\usepackage[normalem]{ulem}
\usepackage[utf8]{inputenc} 
\usepackage{mathtools}

\usepackage{float}
\usepackage{aas_macros}

\usepackage{ulem}
\usepackage{diagbox}

\usepackage[dvipsnames]{xcolor}
\usepackage{mathrsfs}

\newcommand{\be}{\begin{equation}}
\newcommand{\ee}{\end{equation}}
\newcommand{\beqa}{\begin{eqnarray}}
\newcommand{\eeqa}{\end{eqnarray}}

\newcommand\p{{\bm p}}
\renewcommand\k{{\bm k}}
\newcommand\q{\bm{q}}

\newcommand\GG{\Gamma_3}
\newcommand\G{\mathcal{G}_2}

\renewcommand\b{\beta}

\newcommand{\bseq}{\begin{subequations}}
\newcommand{\eseq}{\end{subequations}}

\renewcommand{\ln}{\mathop{\rm ln}\nolimits}

\def\vpsi{{\boldsymbol{\psi}}}

\def\gsim{\raise0.3ex\hbox{$\;>$\kern-0.75em\raise-1.1ex\hbox{$\sim\;$}}}
\def\lsim{\raise0.3ex\hbox{$\;<$\kern-0.75em\raise-1.1ex\hbox{$\sim\;$}}}

\def\beqn#1{\begin{equation}\label{#1}}
\def\eeqn{\end{equation}}

\def\beqa#1{\begin{eqnarray}\label{#1}}
\def\eeqa{\end{eqnarray}}

\def\kmax{{k_\text{max}}}
\def\hMpc{h{\text{Mpc}}^{-1}}
\def\Mpch{h^{-1}{\text{Mpc}}}

\def\Z2{$\mathcal{Z_2}$}

\newcommand {\ignore}[1]{}

\begin{document}

\preprint{MIT-CTP/5722}

\title{The Millennium and Astrid galaxies 
in effective field theory: \\
comparison with
galaxy-halo connection models
at the field level
}

\author{Mikhail M. Ivanov }
\email{ivanov99@mit.edu}
\affiliation{Center for Theoretical Physics, Massachusetts Institute of Technology, 
Cambridge, MA 02139, USA} 
 \affiliation{The NSF AI Institute for Artificial Intelligence and Fundamental Interactions, Cambridge, MA 02139, USA}

\author{Carolina Cuesta-Lazaro}
\email{cuestalz@mit.edu}
 \affiliation{The NSF AI Institute for Artificial Intelligence and Fundamental Interactions, Cambridge, MA 02139, USA}
\affiliation{Department of Physics, Massachusetts Institute of Technology, Cambridge, MA 02139, USA}
\affiliation{Center for Astrophysics $\vert$ Harvard \& Smithsonian, 60 Garden St, Cambridge, MA 02138, USA}

\author{Andrej Obuljen}
\affiliation{Department of Astrophysics, University of Zurich, Winterthurerstrasse 190, 8057 Zurich, Switzerland}

\author{Michael W. Toomey}
\affiliation{Center for Theoretical Physics, Massachusetts Institute of Technology, 
Cambridge, MA 02139, USA}

\author{Yueying Ni}
\affiliation{Center for Astrophysics $\vert$ Harvard \& Smithsonian, 60 Garden St, Cambridge, MA 02138, USA}

\author{Sownak Bose}
\affiliation{Institute for Computational Cosmology, Department of Physics, Durham University, South Road, Durham, DH1 3LE, UK}

\author{Boryana Hadzhiyska}
\affiliation{Miller Institute for Basic Research in Science, University of California, Berkeley, CA, 94720, USA}
\affiliation{Physics Division, Lawrence Berkeley National Laboratory, Berkeley, CA 94720, USA}
\affiliation{Berkeley Center for Cosmological Physics, Department of Physics, University of California, Berkeley, CA 94720, USA}

\author{C\'esar Hern\'andez-Aguayo}
\affiliation{Max-Planck-Institut f\"ur Astrophysik, Karl-Schwarzschild-Str. 1,
D-85748, Garching, Germany}
\affiliation{Excellence Cluster ORIGINS, Boltzmannstrasse 2, D-85748 Garching,
Germany}

\author{Lars Hernquist}
\affiliation{Center for Astrophysics $\vert$ Harvard \& Smithsonian, 60 Garden St, Cambridge, MA 02138, USA}

\author{Rahul Kannan}
\affiliation{Department of Physics and Astronomy, York University, 4700 Keele Street, Toronto, ON M3J 1P3, Canada}

\author{Volker Springel}
\affiliation{Max-Planck-Institut f\"ur Astrophysik, Karl-Schwarzschild-Str. 1, D-85748, Garching, Germany}

\begin{abstract} 
Cosmological analyses of 
redshift space clustering data
are primarily based on 
using 
luminous ``red'' galaxies (LRGs) and 
``blue''
emission line galaxies
(ELGs) to trace underlying 
dark matter. 
Using 
the large 
high-fidelity 
high-resolution 
MillenniumTNG (MTNG) 
and Astrid simulations, we study these galaxies 
with the effective field theory (EFT)-based field
level forward model.
We confirm that both red and blue galaxies 
can be accurately 
modeled with EFT at the field level
and their parameters
match those of the phenomenological 
halo-based models. 
Specifically, we
consider the state of the art Halo Occupation Distribution (HOD) 
and High
Mass Quenched (HMQ) models
for the red and blue galaxies,
respectively. 
Our results explicitly
confirm the validity of the 
halo-based models 
on large scales
beyond the two-point statistics.
In addition, we validate 
the field-level HOD/HMQ-based priors
for EFT full-shape analysis. 
We find that the local bias parameters
of the ELGs are in tension with
the predictions of the
LRG-like HOD models 
and present a simple analytic 
argument explaining 
this phenomenology. 
We also confirm that ELGs 
exhibit weaker non-linear 
redshift-space distortions (``fingers-of-God''), 
suggesting that 
a significant 
fraction of their data 
should be perturbative. 
We find that the
response of EFT parameters 
to galaxy selection 
is sensitive to 
assumptions about
baryonic feedback,
suggesting that  
a detailed
understanding of feedback processes 
is necessary for robust 
predictions of EFT parameters.
Finally, using 
neural density estimation
based on paired HOD-EFT 
parameter samples,
we obtain optimal
HOD models
that reproduce the clustering
of Astrid and MTNG galaxies.
\end{abstract}

\maketitle

\section{Introduction}

Observations of clustering 
of galaxies on large cosmological 
scales is a key source 
of information about 
the constitution, 
expansion history, 
and initial conditions
of our Universe. 
The growing importance 
of galaxy clustering 
is reflected by investments 
into current and future surveys
such as 
DESI~\cite{Aghamousa:2016zmz}, Euclid~\cite{Laureijs:2011gra}, LSST~\cite{LSST:2008ijt}, and Roman~\cite{Akeson:2019biv}.
Realizing the potential 
of these surveys' data requires
an
understanding of main galaxies used
to map our Universe. 
For the current and near-term
future surveys the most 
common types are
red luminous galaxies (LRGs)
and emission line galaxies (ELGs),
often dubbed simply as 
red and blue galaxies, respectively~\cite{eBOSS:2020yzd,DESI:2024uvr}.

The most accurate way to understand 
fundamental cosmological 
properties of galaxies is to rely on
first-principle hydrodynamical (hydro)
simulations, which self-consistently 
solve the coupled
equations for the 
evolution of gas and 
dark matter. Relatively recently hydrodynamical simulations have become large enough to allow for precision studies of large-scale clustering properties relevant for cosmological analyses~\cite{McCarthy:2016mry,Springel:2017tpz,Hernandez-Aguayo:2022xcl}.
A computationally inexpensive alternative to complete hydro simulations 
is given by empirical 
prescriptions to ``paint''
galaxies onto dark matter
halos. Such empirical models
of the galaxy-halo connection,
such as halo occupation distribution (HOD)~\cite{Berlind:2002rn,Zheng:2007zg,Hearin:2015jnf} or 
the subhalo abundance matching (SHAM)~\cite{Conroy:2005aq} (see~\cite{Wechsler:2018pic} for review), 
allow for an efficient statistical 
description
of basic clustering observables.
The HOD models were originally
developed for the description of 
red galaxies, and then 
later extended 
to the blue galaxies in~\cite{Alam:2019pwr,eBOSS:2020yql,Yuan:2021izi}.

On very large, quasi-linear
scales, one can use the perturbative 
bias expansion to model 
the galaxy density field, see~\cite{Desjacques:2016bnm} for a review. 
The perturbative techniques 
have recently experienced
a revival of interest spurred by the
development of the effective field
theory (EFT) of 
large-scale structure~\cite{Baumann:2010tm,Carrasco:2012cv,Ivanov:2022mrd} in
its various formulations~\cite{Blas:2015qsi,Blas:2016sfa,Chen:2020zjt,Chen:2020zjt}, and their successful application in extraction of cosmological
parameters from the galaxy clustering data~\cite{Ivanov:2019pdj,Ivanov:2019hqk,DAmico:2019fhj,Philcox:2021kcw,Chen:2021wdi,Chudaykin:2022nru,Chen:2024vuf}. Although EFT applies only on quasi-linear
scales, it relies only on
symmetries and dimensional 
analysis, and hence is largely
agnostic about the 
galaxy formation physics on small 
scales. EFT captures different 
types of galaxies by means of the 
EFT bias parameters, see~\cite{Desjacques:2016bnm}
for a review. 
The EFT parameters can be predicted
in 
a given model of the galaxy distribution,
such as HOD or the hydro simulation. 
This knowledge can be used 
to inform priors on EFT parameters
in the EFT-based data analyses. 
Early examples of a similar approach 
applied in perturbation theory based 
galaxy clustering analyses
can be found in e.g.~\cite{BOSS:2016psr,Sullivan:2021sof,Kokron:2021faa}.
This idea has been recently
amalgamated into the EFT-based full-shape
analysis with HOD-based priors based on 
neural density estimation,
developed and applied to data in~\cite{Ivanov:2024hgq,Ivanov:2024xgb}, see also~\cite{Cabass:2024wob,Akitsu:2024lyt}.
This approach makes use 
of the field-level EFT 
technique~\cite{Schmittfull:2014tca,Lazeyras:2017hxw,Abidi:2018eyd,Schmidt:2018bkr,Schmittfull:2018yuk,
Elsner:2019rql,
Cabass:2019lqx,Modi:2019qbt,
Schmidt:2020tao,
Schmidt:2020viy,Schmittfull:2020trd,Lazeyras:2021dar,Obuljen:2022cjo,Stadler:2023hea,Nguyen:2024yth,Foreman:2024kzw}, which  
enables one to 
cancel cosmic variance 
and thus achieve
a high precision
of EFT parameter 
measurements
from small simulation volumes.\footnote{Achieving 
a similar precision
with the clustering 
observables, such as
the power spectrum
and bispectrum requires
significantly larger 
simulation volumes, $\sim 600~[h^{-1}\text{Gpc}]^3$~\cite{Ivanov:2021kcd,Ivanov:2021fbu}.}
The HOD-based priors lead to substantial 
improvements 
of the cosmological constraints~\cite{Ivanov:2024hgq,Ivanov:2024xgb}.
The increased statistical power calls
for the validation of the HOD-based priors
with the results of the \textit{ab initio}
galaxy modeling from hydrodynamic
simulations. 

The hydro galaxies 
from the IllustrisTNG simulation~\cite{Springel:2017tpz}
have been studied in the context of the 
field-level EFT models in refs.~\cite{Barreira:2021ukk,Lazeyras:2021dar}.
These works measured relations between the bias
parameters of the hydro galaxies and 
compared them with those of 
dark matter halos. 
In this work 
we build on results of~\cite{Barreira:2021ukk,Lazeyras:2021dar}
and extend them by 
(a) utilizing the large volume high-fidelity MilleniumTNG (MTNG) simulation~\cite{Pakmor:2022yyn} which provides a higher precision than IllustrisTNG, 
(b) comparing the bias parameters of the
hydro galaxies with those of 
the state-of-the-art halo models with assembly bias,
and (c) focusing specifically on the red and blue
galaxies that closely match those observed
in actual surveys such as BOSS~\cite{BOSS:2016wmc} and DESI~\cite{DESI:2024uvr}. 

In addition to MTNG,
we analyze galaxies from
the large volume 
high resolution Astrid simulation~\cite{Bird:2022ulj,2024arXiv240910666N}. 
Astrid uses a 
galaxy formation model 
different 
from MTNG,
with many unique features associated 
with both the low and high redshift physics.
In particular, Astrid has a 
different active galaxy nucleus (AGN) implementation.
Thus, the comparison of MTNG and Astrid 
results will allow us to estimate 
the sensitivity of EFT parameters
to the underlying galaxy formation
scenario.

Another motivation for our study is
to test the accuracy 
of the 
HOD modeling for the red and blue 
hydro galaxies at the field
level. In terms of the actual clustering 
observables, the HOD and hydro galaxies
have been extensively compared in the past at the level
of the two-point function~\cite{Hadzhiyska:2020iri,Hadzhiyska:2022tvd,Hadzhiyska:2022ugr,Bose:2022ibz}
and the squeezed limit of the three-point function~\cite{Yuan:2018qek}. 
In this work, we confront 
the two approaches at the field level, albeit only on large scales
where the two can be connected via 
their EFT parameters.

The halo-based models have been utilized
in various simulation-based
analyses, 
e.g. in a recent community data challenge
``Beyond-2pt''~\cite{Beyond-2pt:2024mqz},
see also~\cite{Valogiannis:2023mxf,Hou:2024blc,Hahn:2023udg,Hahn:2023udg,SimBIG:2023gke,SimBIG:2023ywd,SimBIG:2023nol}.
The success of simulation-based 
techniques in this challenge calls 
for additional tests of
the HOD modeling at the field level. 
Such tests are especially important for 
ELGs, which are less understood
than the red galaxies. 
Even in the context 
of perturbation theory, 
the ELGs have been 
studied
only at the level of the two-point function
and only for HOD-based simulations~\cite{Mohammad:2017lzz,eBOSS:2020fvk,Ivanov:2021zmi,DESI:2024jis}.
Importantly, these analyses 
showed that these galaxies
feature a weaker fingers-of-God signature
than the LRGs, which could extend the 
validity of the perturbative regime for them and hence provide 
tighter 
cosmological constraints~\cite{Chudaykin:2019ock,Sailer:2021yzm,Ivanov:2021zmi,Ivanov:2021fbu,Ivanov:2021fbu}. 
This observation calls for a further 
validation against the high
quality hydrodynamic galaxies,
which we present here.

From the galaxy-halo connection side, we use the baseline modified High Mass Quenched (HMQ) model 
for ELGs adopted by DESI~\cite{DESI:2023ujh,DESI:2024rkp}, which is built on~\cite{Alam:2019pwr,eBOSS:2020yql,Yuan:2021izi,Rocher:2023zyh}.
We create samples of EFT parameters
following the same approach as~\cite{Ivanov:2024hgq,Ivanov:2024xgb},
and compare them against 
the measurements from blue galaxies of 
MTNG and Astrid simulations.

Our paper is structured as follows. We present the main
results in Section~\ref{sec:summ}.
Our methodology, EFT and HOD modeling, and data 
are discussed in Section~\ref{sec:method}.
Section~\ref{sec:long_results} describes 
our analysis and results in detail. 
There we compare the EFT parameters
of hydrodynamic galaxies 
with the HOD-based priors. 
Section~\ref{sec:disc} draws
conclusions.

\begin{figure*}
\centering
\includegraphics[width=0.99\textwidth]{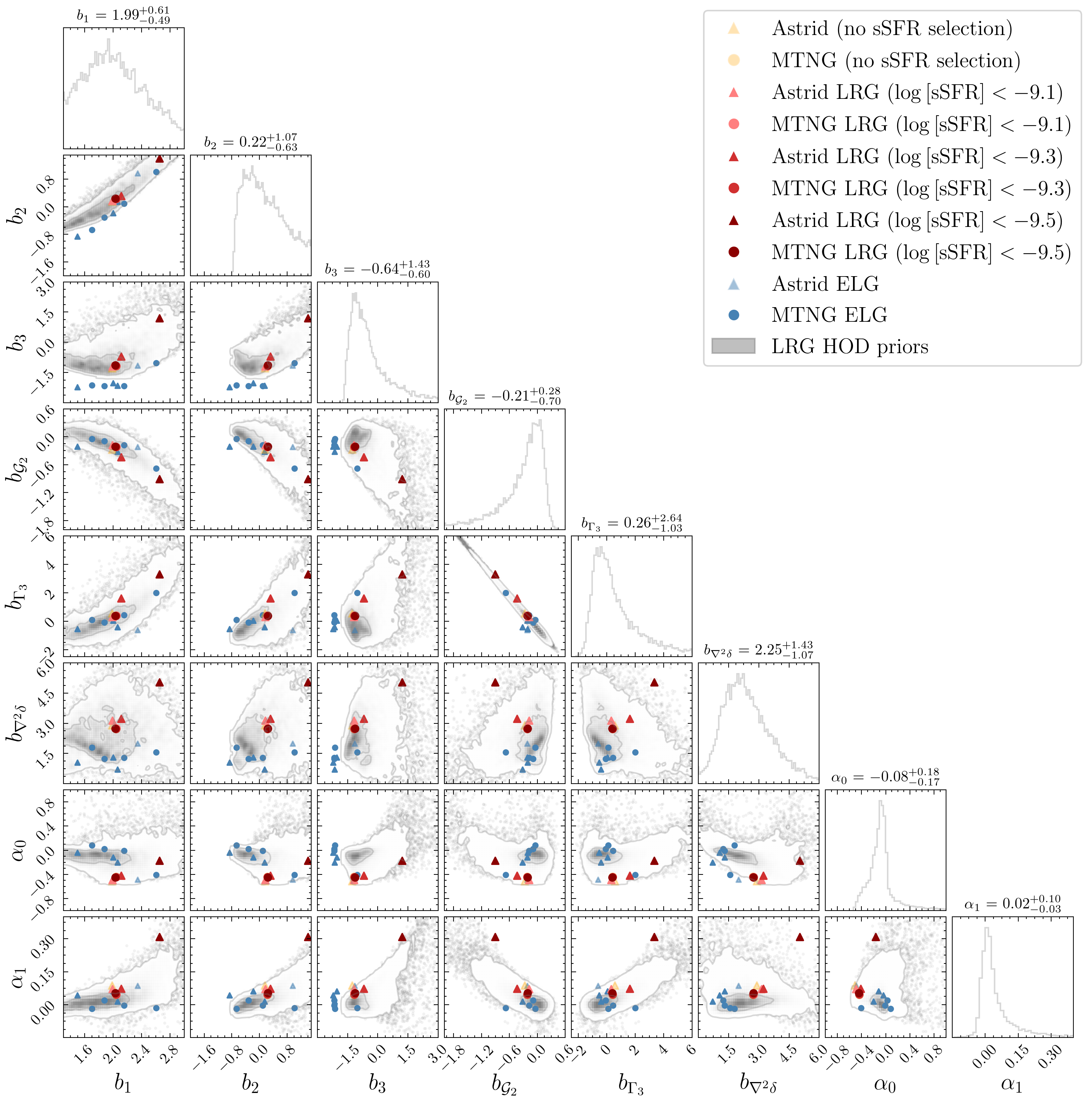}
   \caption{MTNG and Astrid bias parameters for LRGs against the LRG-HOD-based density distribution (LRG-HOD priors). For comparison, we also show four ELGs samples. 
    } \label{fig:dist_bias}
\end{figure*}

\begin{figure*}
\centering
\includegraphics[width=0.99\textwidth]{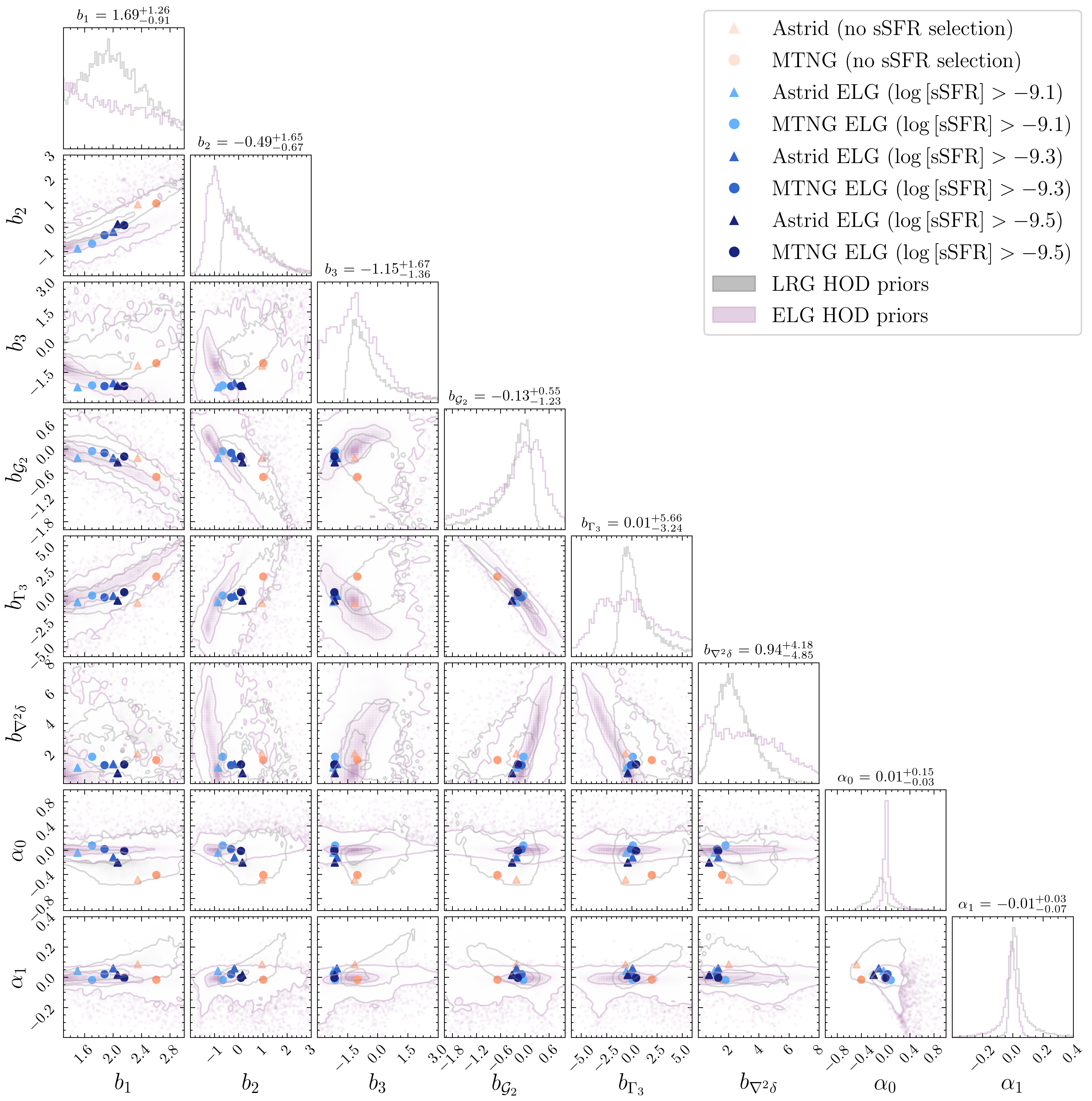}
   \caption{MTNG and ASTRID bias parameters of ELGs vs. the distributions extracted from LRG-based HOD and ELG-based HMQ catalogs.
    } \label{fig:bias_elg}
\end{figure*}

\section{Main results}
\label{sec:summ}

Let us start with the summary of our 
methodology and main
results. 
We extract EFT parameters
of BOSS and DESI-like LRGs at $z=0.5$
and ELGs at $z=1$
from MTNG and Astrid simulations. 
We consider several different LRG and 
ELG
selections.
Specifically, following~\cite{Hadzhiyska:2020iri,Bose:2022ibz,Hadzhiyska:2022tvd,Hadzhiyska:2022ugr},
we use the cuts on the number density (equivalent to the stellar mass) 
and the specific star formation
rate (sSFR),
i.e.\ the star formation rate normalized by the galaxy’s stellar mass.
We use $h\cdot$yr$^{-1}$
units 
for sSFR. 
Physically sSFR$^{-1}$
is the typical timescale
over which the 
stellar mass doubles.

We measure EFT parameters
by fitting the 
transfer 
functions of the 
field-level model of~\cite{Schmittfull:2018yuk,Schmittfull:2020trd}
following the approach 
of~\cite{Ivanov:2024hgq,Ivanov:2024xgb}.\footnote{These are produced with the code \texttt{Hi-Fi mocks} publicly available at 
\url{https://github.com/andrejobuljen/Hi-Fi_mocks}.
}
Our measurements 
of the real-space 
EFT parameters
for DESI-like LRGs and ELGs
are displayed in figs.~\ref{fig:dist_bias},~\ref{fig:bias_elg} 
respectively.
Our main findings are:

\textbf{1. Consistency of hydro-based EFT parameters with HOD values for LRGs.}
We have found all EFT parameter 
values for LRGs are highly consistent 
with the HOD-based values from the decorated models 
considered in refs.~\cite{Ivanov:2024hgq,Ivanov:2024xgb},
see fig.~\ref{fig:dist_bias}.
This confirms that the HOD priors
from these works are robust under
the variation of the underlying 
galaxy distribution model. 
Our results also explicitly
confirm the validity 
of the LRG-HOD models 
beyond the two point function statistic,
albeit only on quasilinear scales.

We have also confirmed 
previous reports~\cite{Eggemeier:2021cam,Ivanov:2021kcd,Ivanov:2024xgb} that the 
quadratic bias parameters $b_2$
for LRG-HOD 
are larger than those
of the dark matter halos
~\cite{Lazeyras:2015lgp}
for $b_1\sim 2$.
We also show that this behavior
naturally follows from the 
shape of the LRG-HOD function
and the dark matter halo mass
function. 

\textbf{2. Modeling ELGs at  
the field level.}
We have carried out precision measurements of the full set of EFT 
parameters for ELGs at the field level for the first time.
Our results agree with previous 
analyses based on the eBOSS ELG data~\cite{eBOSS:2020fvk,Ivanov:2021zmi} 
and the early ELG-halo 
based models~\cite{Alam:2019pwr,eBOSS:2020xlj}. 
Our measurements 
of EFT parameters 
can be used to inform 
and validate 
full-shape analyses 
of the DESI ELGs~\cite{DESI:2024jis}.

\textbf{3. Local bias 
parameters of ELGs.}
One interesting outcome of our measurements 
is that ELG's 
combinations of local (in matter density) bias parameters
$\{b_1,b_2,b_3\}$ appear to be in a very 
strong tension with the HOD
values when using 
the LRG-based halo models. There is no 
(decorated) HOD model for LRGs
that can reproduce the combination of large-scale
local bias parameters of ELGs.
This is expected 
from the phenomenological
point of view 
as the blue galaxies quench 
at late stages of their evolution 
(after undergoing a few mergers) 
and no longer produce strong emission lines.

In contrast
to the local bias parameters, the other EFT parameters of ELGs beyond the local ones
appear very similar to the 
LRG-HOD
ones. 
This means that one can,
in principle, 
reproduce non-local EFT parameters 
of ELGs with an appropriate choice
of parameters of the
LRG-HOD models. 
The non-LRG-like 
distribution
of the local 
bias coefficients 
is an interesting signature of 
ELGs, which we show
to directly follow
from their halo occupation shape. 

In contrast to the 
$\{b_1,b_2\}$ values
of LRG-HOD galaxies, 
the local 
ELG bias parameters that we have measured
are consistent 
with 
those of the underlying 
dark matter halos. 
We present a simple 
analytic argument 
explaining this phenomenology. 
It will be interesting to  develop 
more sophisticated 
semi-analytic models of ELGs along the lines of 
\cite{Marinucci:2023jag}.

\textbf{4. Consistency of HMQ models with hydro galaxies at the field level.}
To extend the
simulation based prior approach to 
ELGs we have generated 
a large sample of EFT parameters
using 10500 synthetic 
catalogs based on the phenomenological
HMQ model. We have shown that
the HMQ model captures 
well the above ``anomalous''
behavior of the local bias
parameters, see fig.~\ref{fig:bias_elg}.
It also correctly
reproduces the weaker 
fingers-of-God signature
observed for the ELGs. 
Therefore, our 
analysis 
confirms the validity 
of the HMQ model at the field level (albeit only on large scales). 
Thus, our work 
supports
validation studies of the HMQ model
based on 
the two-point functions (e.g.~\cite{DESI:2023ujh})
and extends them
to higher-order 
clustering observables.

\textbf{5.  Weak non-linear redshift-space distortions of ELGs.}
We have found that ELG from the hydrodynamic simulations feature
weaker non-linear redshift-space
distortions, known as the 
``fingers-of-God'' (FOG)~\cite{Jackson:2008yv}. 
This signature
is well reproduced by the HMQ 
models: the 
distribution of the EFT 
redshift-space
counterterms  
from these models is significantly 
narrower than that of LRG-HOD.
If this is an actual physical property 
of the ELG in our Universe, 
it can have a 
profound 
impact on the perturbative 
analyses of ELG samples.
Weaker FOG imply a larger cutoff scale 
of EFT, thereby increasing $\kmax$
in cosmological analyses and yielding
more data in the perturbative regime~\cite{Chudaykin:2019ock,Sailer:2021yzm,Ivanov:2021zmi,Ivanov:2021fbu}. 

From the physical perspective, weaker FOG 
follow from the fact
that ELGs reside
in small mass host halos
which have a relatively low 
(i.e. ``cold'') 
velocity 
dispersion. 

\textbf{6. The response of the EFT parameters to the specific star formation rate.}
We have studied the dependence of the EFT parameters on galaxy selection.
We find that the response 
of Astrid and MTNG LRGs 
to sSFR is quite different. 
The EFT parameters of MTNG LRGs respond 
to sSFR very weakly\footnote{It is consistent with results for IllustrisTNG~\cite{Barreira:2021ukk}.},
while for Astrid we observe
some noticeable response whose 
strength
grows for smaller sSFR cuts. 
In terms of the EFT
parameter values,
the results of Astrid and MTNG 
are broadly consistent for 
LRGs selected 
with sSFR thresholds 
corresponding to  
fast star formation.
However, Astrid and MTNG 
predict quite different 
LRGs if we apply 
more realistic 
sSFR cuts
consistent 
with slow star formation.
In this case Astrid galaxies live 
in more massive halos, 
which results in larger values 
of EFT bias parameters. 
We argue that the discrepancy
between the LRG sSFR responses 
can be explained by 
the difference in the 
MTNG and Astrid 
feedback models.
Specifically, in the context of the sSFR sensitivity, 
the most relevant feedback parameters
are AGN response, 
gas cooling, star formation, and quenching mechanisms.

For ELGs, Astrid and MTNG galaxies
are quite consistent
as functions of sSFR, 
and they both
predict a strong dependence on 
sSFR thresholds. The exact form 
of the response
is however slightly 
different, which again, 
can be traced back to
the differences in
feedback.

Finally, let us note that 
the simulation-based prior approach 
proposed in~\cite{Ivanov:2024hgq,Ivanov:2024xgb}
can be applied to hydrodynamic simulations
as well. This will require 
producing a suite of simulations with different 
subgrid models, similar 
to CAMELS~\cite{CAMELS:2020cof},
but in a cosmologically large
volume. 

\textbf{7. Optimal galaxy-halo connection for hydrodynamic galaxies via EFT parameters.}
The HOD priors 
for EFT parameters
from~\cite{Ivanov:2024hgq,Ivanov:2024xgb} can be used to map the bias parameters onto 
HOD models via a conditional 
distribution modeled with normalizing 
flows. Then, the EFT parameters of hydrodynamic
galaxies can be used to predict optimal
values of HOD parameters that are needed 
to reproduce the large-scale clustering 
of hydro galaxies within the HOD 
framework. This approach is based only
on using the large-scale clustering 
information, but it allows for 
an efficient, cheap, and accurate matching of the clustering 
observables at the field level. 
This approach may be useful for
calibration of HOD models at the field
level. 

We realize this idea in practice 
and explicitly 
present HOD/HMQ models needed to reproduce
the clustering of hydro galaxies
with the target EFT parameters from Astrid and MTNG.
The optimal 
halo model parameters we find
are consistent with halos
hosting the hydro galaxies.

\section{Methodology}
\label{sec:method}

The main goal of this work is to measure
EFT parameters of the BOSS-like and 
DESI-like hydro galaxies
and compare them with those of the 
HOD/HMQ based galaxies. To extract the EFT parameters at high precision, we use
the field-level EFT forward model 
which allows for cosmic variance cancellation. 
Let us describe now the simulations that we use here, 
the galaxy selection, 
the HOD models, and the field-level EFT 
technique. 

\subsection{Hydrodynamic Simulations}

We start with a brief description of the 
hydrodynamical simulations studied
in this work. 

\textbf{MTNG.} 
We use the largest hydrodynamic box  
MTNG740 with $500\,\Mpch$ side length. 
This box contains 
$2 \times 4230^3$ resolution elements.
MTNG is based on the 
underlying cosmology of the IllustrisTNG simulation: 
$\{\Omega_m,\Omega_b,h,\sigma_8,n_s,M_\nu/\text{eV}\}=$
$\{0.3089,0.0486,0.6774,0.8159,0.9667,0\}$, 
where $\Omega_m,\Omega_b$
are late-time abundances
of matter and baryons, 
$h$ is the reduced
Hubble constant, 
$\sigma_8$
is the mass fluctuation
amplitude, $n_s$
is the primordial
spectral tilt,
and $M_\nu$
is the total neutrino 
mass. 

The key feature of 
MTNG that is crucial for our 
study is the large volume,
providing small
statistical errors, 
which is crucial for 
accurate measurements
of the large-scale EFT parameters. 
The MTNG simulation is based on the 
physics model
detailed in
\cite{2017MNRAS.465.3291W,Pillepich:2017jle,Nelson:2017cxy,2018MNRAS.477.1206N,2018MNRAS.480.5113M,Springel:2017tpz,Nelson:2019jkf,2019MNRAS.490.3196P}.
We refer the reader to
\cite{Hernandez-Aguayo:2022xcl,Pakmor:2022yyn}
for further technical details
and outputs of the MTNG simulation.

\textbf{Astrid.}
In addition to MTNG, we also include another cosmological galaxy formation simulation Astrid \cite{Bird:2022ulj} as a complementary study. 
Astrid is conducted within a 250 $\Mpch$ box with $2 \times 5500^3$ particles, achieving a baryon mass resolution of $\sim 10^6 M_{\odot}$. 
This allows Astrid to resolve galaxies and halos down to lower mass scales, though it introduces larger statistical scatter on large scales. 
The cosmological parameter employed by Astrid are based on \cite{Planck}:  
$\{\Omega_m,\Omega_b,h,\sigma_8,n_s,M_\nu/\text{eV}\}=$
$\{0.3089,0.0486,0.6774,0.816,0.9667,0.06\}$.

The Astrid galaxy formation model is described in detail in~\cite{Bird:2022ulj}. 
In particular, gas cools via primordial radiative cooling \cite{Katz1996} and metal-line cooling, with gas and stellar metallicities evolving according to the prescriptions in \cite{Vogelsberger2014}. 
The simulation adopts patchy reionization \cite{Battaglia2013} and employs the ionizing UV background from \cite{FG2020}, along with gas self-shielding \cite{Rahmati2013}. 
Star formation follows a multi-phase model \cite{SH03} that accounts for the influence of molecular hydrogen \cite{Krumholz2011}. 
Feedback from Type II supernovae is implemented following \cite{Okamoto2010}, with wind speeds proportional to the local dark matter velocity dispersion.
Black hole seeding, accretion, feedback, and dynamics are modeled as described in \cite{Ni2022}.

\subsection{Galaxy populations}
\label{sec:gal_pops}

\begin{figure*}[htb]
\centering
\includegraphics[width=0.99\textwidth]{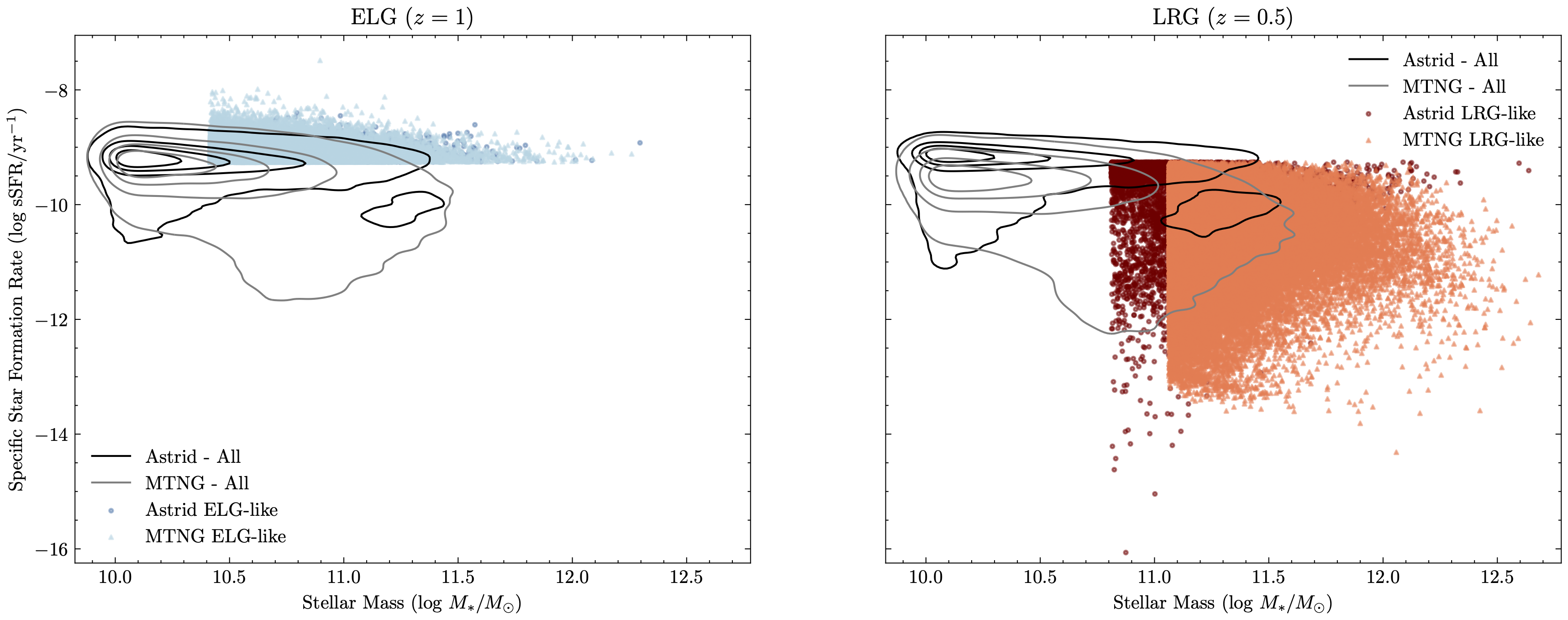}
   \caption{The samples of MTNG 
   and Astrid galaxies 
   in the stellar mass - specific 
   star formation rate planes. 
   We display blue ELGs ($z=1$) on the left panel
   and LRGs ($z=0.5$) on the right panel. The colored distributions display
   the selection with the threshold 
   $-\log(\text{sSFR/$h$yr}^{-1})=9.0862$
   and stellar mass cuts 
   adjusted to 
   reproduce 
the target number
   densities 
   described in the main text. 
   The black and gray contours 
   contain 68\% and 95\% of
   all galaxies for Astrid and MTNG 
   simulations, respectively.
    } \label{fig:selection}
\end{figure*}

\textbf{MTNG.} 
We select samples that 
are similar to 
BOSS LRGs, DESI LRGs, 
DESI ELGs catalogs. 
The redshifts of the samples
are $z=0.5$ for LRGs and
$z=1$ for ELGs. 
ELGs and LRGs are selected 
by applying specific
star formation cuts following~\cite{Hadzhiyska:2020iri,Bose:2022ibz,Hadzhiyska:2022tvd,Hadzhiyska:2022ugr},
at a fixed target number density.
The SFR cut selection provides 
a good approximation to samples of [OII] emitting galaxies.
A more robust
way to identify 
emission line 
galaxies is to model the colors and emission lines in detail with a population synthesis model. 
However, as shown in~\cite{Hadzhiyska:2020iri}, a careful color modeling with population synthesis and matching to color-color data 
consistent with the DESI observations produces results similar to the ones coming
from the sSFR-selection.

In our selection, the total stellar mass is adjusted 
to reproduce the target number density,
which is 
equivalent to applying 
stellar mass cuts as  in~\cite{Hadzhiyska:2020iri}. 
The BOSS/DESI samples
have number densities 
$3.5\times 10^{-4}$ $[\Mpch]^{-3}$
and $5\times 10^{-4}$ $[\Mpch]^{-3}$, 
respectively. 
All ELG samples have the DESI-like number density
$5\times 10^{-4}~[h^{-1}\text{Mpc}]^{-3}$~\cite{DESI:2024uvr,DESI:2023ujh}.

For each
number density choice 
we consider four different sub-samples:
all galaxies, and 
three samples whose 
specific 
star formation rate (sSFR)
is lower than a fixed threshold. 
Physically, sSFR$^{-1}$
is the typical timescale over which 
the stellar mass in the galaxy doubles
due to active star formation.
It is convenient to use the variable 
$\tau_{\rm sSFR}=-\log_{10}(\text{sSFR}/(h\cdot\text{yr}^{-1}))$, i.e.\ 
the logarithm of 
this time scale. 
Larger values of $\tau_{\rm sSFR}$
imply slower star mass accumulation. 
The sSFR thresholds for LRGs 
we consider are
$\tau_{\rm sSFR}=9.0862$,
$\tau_{\rm sSFR}=9.2596$,
and 
$\tau_{\rm sSFR}=9.5376$.
Our selection for one fixed sSFR cut
is illustrated in Fig.~\ref{fig:selection}.
Note that ``all galaxies'' 
implies no sSFR cut whatsoever, 
and hence this sample 
contains both LRGs and ELGs.
Fig.~\ref{fig:selection} shows
that this sample is, however, 
largely dominated by LRGs. 
Also note that~\cite{Hadzhiyska:2022ugr}
used $\tau_{\rm sSFR}=9.0862$
as a baseline choice.

Following~\cite{Hadzhiyska:2022ugr}, ELGs are 
selected by applying the 
specific 
star formation rate cut $\tau_{\rm sSFR}=9.0862$. This should be 
treated as a baseline choice 
of ELG selection. 
In addition, we consider two 
higher $\tau_{\rm sSFR}$, i.e. 
smaller star formation rates, 
$\tau_{\rm sSFR}=9.2596$,
and 
$\tau_{\rm sSFR}=9.5376$,
and also a sample of all galaxies. 
Note that we do not expect that the latter three 
samples match the DESI galaxies well,
especially the ``All'' samples
that do not apply any sSFR cut,
and which are actually dominated by LRGs.
Nevertheless, 
bracketing 
the sSFR values 
provides an insight into the 
sensitivity 
of the ELG clustering to the applied sSFR cuts.

\textbf{Astrid.}
For LRGs and ELGs, we implement 
the selection similar to the MTNG case:
we consider
the samples with 
different $\tau_{\rm sSFR}= 9.0862,9.2596,9.5376$, 
and all galaxies,
at fixed number densities. 
The redshifts are $z=0.5$ (LRGs)
and $z=1$ (ELGs).

Note that in both MTNG and Astrid, AGN feedback powered by central supermassive black holes (SMBHs) is the primary quenching mechanism for massive galaxies, leading to the rising of quiescent (red) galaxy population and account for the color bimodality of galaxies in the low-redshift universe. 
Although the AGN feedback prescriptions in Astrid and TNG are broadly similar—employing both thermal and kinetic feedback to mimic the quasar and jet modes of AGN activity—Astrid's AGN feedback is comparatively milder.\footnote{See~\cite{Pandey:2023wqp} for a comparison of TNG and Astrid 
baryonic 
feedbacks
in the context of the weak lensing surveys. Note that
from the weak lensing perspective
both TNG and Astrid are characterized by 
relatively weak
feedback.} 
Specifically, the jet-mode AGN feedback (which can efficiently quench galaxies) in Astrid activates only for more massive black holes ($M_{\rm BH} \gtrsim 5\times 10^8M_\odot$), whereas in MTNG, this mode turns on at
$M_{\rm BH} \gtrsim 1\times 10^8M_\odot$. 
Consequently, massive galaxies ($M_* \gtrsim 10^{11} M_{\odot}$) in MTNG tend to have lower sSFR than those in Astrid, 
as seen in the galaxy selection plot in fig.~\ref{fig:selection}. 
Most MTNG massive galaxies are strongly quenched 
with very low SFRs, making the LRG selection in MTNG relatively insensitive to the applied sSFR thresholds. 
In Astrid, many massive galaxies have sSFR values close to the  threshold, making the LRG selection quite sensitive to the applied threshold. 

The same argument 
predicts a difference 
in clustering properties of
ELGs. Since Astrid 
galaxies are more star forming, 
we expect to find 
a larger abundance of 
ELGs at $z=1$, 
and hence generically they should be 
less biased than MTNG ELGs for the 
same sSFR threshold. 
Both MTNG's and Astrid's densities
of ELG samples seen in 
fig.~\ref{fig:selection}
have a strong response to sSFR, 
but the MTNG response is more
monotonous within the 
sSFR cuts considered here. 
We will confirm both of these
predictions
at the level of the EFT parameters. 

\subsection{Halo occupation models}

One of the goals of our study 
is to test how well the MTNG and Astrid 
galaxies can be described with 
the phenomenological
HOD models. To that end we will
compare EFT parameters 
of hydro galaxies with 
those of the HOD models. 
Let us describe now the HOD models
and priors that we use. 

Following~\cite{Ivanov:2024hgq,Ivanov:2024xgb}, all  HOD samples are 
produced with the 
the \texttt{AbacusSummit small} (covariance) suite~\citep{Maksimova:2021ynf}, with box site length 
500~$\Mpch$, identical 
to that of MTNG.

Our comparison will be 
based on three halo-based galaxy samples.
For LRGs, we consider two decorated 
models: with and without the assembly bias
based on the halo concentration~\cite{Hearin:2015jnf}. 
For ELGs, we consider
the HMQ model plus additional 
assembly bias parameters that capture the
dependence on both the local overdensity
and the halo concentration. 
The LRG catalogs we use were generated 
before in~\cite{Ivanov:2024hgq,Ivanov:2024xgb}.
The HMQ-based catalogs for ELGs
are generated here for the first time.

For the LRGs we start with
the ``standard'' HOD model 
in which the mean HOD 
of centrals and satellites 
are given by~\cite{Zheng:2007zg} 
\be
\label{eq:HODforLRG}
\begin{split}
& \langle N_c\rangle(M)=\frac{1}{2}\left[1+\text{Erf}\left(\frac{\log M-\log M_{\rm cut}}{\sqrt{2}\sigma}\right)\right]\,,\\
& \langle N_s\rangle(M)=\langle N_c\rangle(M)\left(\frac{M-\kappa M_{\rm cut}}{M_1}\right)^{\alpha}\,,
\end{split} 
\ee
where $\text{Erf}(x)$ stands for the Gauss error function,
$M_{\rm cut}$, 
$M_1$,  
$\kappa M_{\rm cut}$,
and $\alpha$ are
free HOD parameters.
To include assembly bias, 
we promote $M_{\rm cut}$
and $M_{1}$ to be functions
of the halo concentration $c$
and the local overdensity $\delta$,
which we rank within
halo mass bins and normalize
the ranks to the range $[0,1]$,
\be 
\label{eq:assembl}
\begin{split}
&\log_{10}M_{\rm cut}
\to \log_{10}M_{\rm cut}\\
&+A_{\rm cent}(c^{\rm rank}-0.5)
+B_{\rm cent}(\delta^{\rm rank}-0.5)\\
& \log_{10}M_{1}
\to \log_{10}M_{1}\\
&+A_{\rm sat}(c^{\rm rank}-0.5)
+B_{\rm sat}(\delta^{\rm rank}-0.5)\,.
\end{split}
\ee 
The concentration is given 
by $c=r_{90}/r_{25}$,
where $r_f$ denotes the radius
enclosing a fraction $f$
of the total halo mass. 
The local overdensity is
defined as
\be 
\delta =\frac{M(r_{98}<r<r_{c})}{\langle M(r_{98}<r<r_{c})\rangle}-1\,,
\ee 
where $M(r)$ is the enclosed mass
of the neighbouring halos
and $r_c=5~\Mpch$.

In addition to the above HOD implementation, we consider 
a satellite radial distance parameter
$s$ which is motivated by the baryonic
feedback~\cite{Yuan:2018qek}. 
To implement it, we
rank all particles inside a halo by their radial distance from the halo
center, and then assign each particle a 
rank-dependent weight. The probability for the i-th ranked
particle to host a satellite galaxy is given by 
\be 
p_i = \frac{\langle N_s\rangle(M)}{N_P}
\left(1+s\left(1-\frac{2r_i}{N_P-1}\right)\right)\,,
\ee 
where $N_P$ denotes the number
of particles in a halo
and $r_i$ is the rank of
particle $i\in \{0,...,N_P-1\}$.

By default the 
velocity of the central galaxy 
is equal to the bulk
velocity of the largest subhalo. 
The velocities of  
satellite galaxies
are matched to the velocities
of their host dark matter
particles in the simulations. 
Following~\cite{Guo:2014iga,Yuan:2021izi}, we also include 
velocity bias. This is done in two steps. To that end we 
scatter 
the line-of-sight
projection of the central galaxies'
velocity with random Gaussian fluctuations $\delta v$ whose variance is 
given by the velocity
dispersion
of dark matter particles
in a given halo $\sigma_{v}$. 
This fluctuation is additionally
rescaled by the velocity bias
parameter $\alpha_c$:
\be 
\begin{split}
& v_{\rm cen, z}=v_{L2,z}+\alpha_c \delta v(\sigma_{v})\,,
\end{split}
\ee 
where $v_{L2,z}$ is the 
line-of-sight projection 
of the subhalo velocity.
As for the satellites, their velocity bias shifts them w.r.t the 
host particle velocity $v_{p}$ as
\be 
\begin{split}
& v_{\rm sat, z}=v_{L2,z}+\alpha_s (v_{p,z}-v_{L2,z})\,,
\end{split}
\ee 
where $\alpha_s$ is the 
satellite velocity bias. 
In the absence of the satellite velocity bias $\alpha_s=1$ and hence 
$ v_{\rm sat, z}=v_{p,z}$.

The first sample of LRG HOD models 
we consider was discussed in~\cite{Ivanov:2024hgq}. It is generated
for the following set of HOD parameters
drawn from uniform priors defined in~\cite{Paillas:2023cpk} (here and in what follows the 
halo 
mass units are $h^{-1}M_\odot$):
\be 
\label{eq:lrghod1}
\begin{split}
& \log_{10} M_{\rm cut}\in [12.4,13.3]\,,\quad 
 \log_{10}  M_1\in [13.2,14.4]\,,\\
& \log_{10} \sigma \in [-3.0,0.0]\,,\quad  \alpha \in [0.7,1.5]\,,\quad \kappa \in [0.0,1.5]\,,\\
& B_{\rm cen} \in [-0.5,0.5]\,,\quad 
B_{\rm sat} \in  [-1.0,1.0]\,.
\end{split}
\ee 
All other parameters are set to zero. 
This constitutes the main
LRG-HOD sample
used in our work. 

The second set of 
catalogs is produced as part
of~\cite{Ivanov:2024xgb}.
It is generated from LRG HOD 
parameters which we randomly sample from
the following flat distribution as suggested in~\cite{DESI:2023ujh}:
\be 
\begin{split}
& 
\log_{10} M_{\rm cut}\in [12,14]\,,\quad 
\log_{10} M_1\in [13,15]\,,\\
& \log \sigma \in [-3.5,1.0]\,,\quad  \alpha \in [0.5,1.5]\,,\quad \\
& \alpha_c \in [0,1]\,,
\quad 
\alpha_s \in [0,2]\,,
\quad s\in [0,1]\,,
\quad  \kappa \in [0.0,1.5]\,,\\
& A_{\rm cen} \in [-1,1]\,,\quad 
A_{\rm sat} \in  [-1,1]\,,\\
& B_{\rm cen} \in [-1,1]\,,\quad 
B_{\rm sat} \in  [-1,1]\,.
\end{split}
\ee 
In what follows we call this the LRG-HOD-II sample, as opposed 
to the LRG-HOD-I sample specified by eq.~\eqref{eq:lrghod1}.
The comparison between the two LRG 
catalogs can be used to estimate
the impact of additional parameters.
In addition, the RSD is consistently
treated only in the second 
catalog. Our baseline redshift
for both LRG HOD catalogs is $z=0.5$. 

Note that we use the CompaSO
halo finder to generate the HOD priors~\cite{Hadzhiyska:2021zbd}.
In principle, the HOD priors 
are expected to depend
on halo finding, and hence the halo finding
mechanism should be treated 
as an extra ``microscopic'' parameter. 
The results of~\cite{Ivanov:2024xgb,Akitsu:2024lyt},
however, imply that the dependence 
of the EFT parameter priors
(especially those capturing  
galaxy bias)  
on the halo finder is quite weak in practice.\footnote{The discrepancy between
the friends-of-friends (FoF)~\cite{1985ApJ...292..371D}
and Rockstar methods is 
quite significant 
at the level of the redshift-space counterterms. However, the standard FoF 
does not consistently incorporate velocity information,
and hence Rockstar is expected
to provide more accurate results. 
}
In particular, 
CompaSO and Rockstar~\cite{2013ApJ...762..109B} produce 
indistinguishable results
for the quadratic bias parameters~\cite{Ivanov:2024xgb,Akitsu:2024lyt}. CompaSO is very similar to Rockstar
in terms of velocity information~\cite{Hadzhiyska:2021zbd},
and hence the distribution of redshift space EFT parameters from the two 
halo finders
is also expected to be largely similar. 
Therefore, we will ignore
the sensitivity to halo finding
in what follows. 

For the ELGs, we use the baseline
HMQ model, 
in which the mean HOD of centrals
is given by~\cite{DESI:2023ujh}
\be 
\begin{split}
\langle N_c\rangle &=2A\phi(M)\Phi(\gamma M)\\
&+\frac{1}{2Q}\left(1+\text{Erf}\left(\frac{\log(M/M_{\rm cut})}{0.01}\right)\right)\,,
\end{split}
\ee 
where we set the quenching efficiency to 
$Q=100$ following~\cite{Yuan:2022rsc} 
and other functions are 
defined as
\be
\begin{split}
&
\phi(x) = \mathcal{N}(\log M_{\rm cut},\sigma_M)\,,\\
&\Phi(x) = \frac{1}{2}\left[1+\text{Erf}\left(\frac{x}{\sqrt{2}}\right)\right]\,,\\
& A=\frac{p_{\rm max-1/Q}}{\max[2\phi(x)\Phi(\gamma x)]}\,.
\end{split}
\ee 
For the satellites, we use the same 
power-law HOD as for LRGs, eq.~\eqref{eq:HODforLRG}.
In addition, we include assembly bias
and velocity bias 
using the same approach as for LRGs, i.e. 
eqs.~\eqref{eq:assembl}. We sample
the HMQ parameters from the following
uniform distributions: 
\be 
\begin{split}
& 
\log M_{\rm cut}\in [11.6,12.6]\,,\quad 
\log M_1\in [12.5,18]\,,\\
& \log \sigma \in [-3.5,1.0]\,,\quad  \alpha \in [0,1.2]\,,\quad \\
& 
\alpha_c \in [0,1]\,,
\quad 
\alpha_s \in [0,2]\,,
\quad s\in [-1,1]\,,
\quad  \kappa \in [0,10]\,,\\
& p_{\rm max}\in [0.05,1]\,,\quad \gamma \in [1,15] \\
& A_{\rm cen} \in [-1,1]\,,\quad 
A_{\rm sat} \in  [-1,1]\,,\\
& B_{\rm cen} \in [-1,1]\,,\quad 
B_{\rm sat} \in  [-1,1]\,.
\end{split}
\label{eq:elg-prior}
\ee 
We do no vary $Q$ following~\cite{DESI:2023ujh},
but in principle one can 
easily sample this parameter too. 

We generate 10500 catalogs 
of ELG-type galaxies 
based on the \texttt{AbacusSummit small}
suite. Our baseline redshift
for ELGs is $z=1.1$.
To obtain the EFT parameters,
we fit the synthetic galaxy 
density maps in redshift space
with the field-level EFT forward 
model of~\cite{Schmittfull:2018yuk,Schmittfull:2020trd,Obuljen:2022cjo,Ivanov:2024hgq,Ivanov:2024xgb}.
Our approach is identical 
to the one used in~\cite{Ivanov:2024hgq,Ivanov:2024xgb}.
The details of the
field-level forward model are given below.

\subsection{Field level EFT}

Our EFT parameter fits 
are based
on the field-level forward model
introduced in refs.~\cite{Schmittfull:2018yuk,Schmittfull:2020trd}. 
Its key elements are
various shifted operators
in the bias expansion. 
The shifted operators are defined 
as
\be 
\label{eq:shift}
\tilde{\mathcal{O}}(\k)
=\int d^3 \q~\mathcal{O}(\q)
e^{-i\k\cdot(\q+\vpsi_1(\q)+f\hat{\bm{z}}(\vpsi(\q)\cdot \hat{\bm{z}}))}~\,,
\ee 
where $\q$ is the Lagrangian coordinate, $\mathcal{O}(\q)$
is an operator in Lagrangian space, 
$\vpsi_1$ is the Zel'dovich displacement, 
\be 
 \bm{\psi}_1(\q,z)=\int d^3q~e^{i\q\k}\frac{i \bm{k} }{k^2}\delta_1(\k,z)\,,
\ee 
and $f$ is the logarithmic growth rate. $\delta_1$ above 
is the linear matter density field,
which obeys the Gaussian statistic:
\be 
\langle \delta_1(\k)\delta_1(\k')\rangle =(2\pi)^3\delta_D^{(3)}(\k'+\k)P_{11}(k)\,.
\ee 
In what follows we will use 
primes to denote stripping off 
the Dirac delta function,
\be 
\langle \delta_1(\k)\delta_1(\k')\rangle'=P_{11}(k)\,.
\ee 
For real space (rest-frame)
observables we set $f=0$ in 
eq.~\eqref{eq:shift}. 
To avoid operator mixing, the shifted 
operators are orthogonolized as
\be 
\langle \tilde{\mathcal{O}}^\perp_m 
\tilde{\mathcal{O}}^\perp_n \rangle  = 0\quad \text{if}
\quad n\neq m\,.
\ee 
The EFT forward model 
in redshift space is given by
\be 
\label{eq:eft-field_rsd}
\begin{split}
&\delta^{\rm EFT}_g (\k,\hat{\bm z})= \delta_Z(\k,\hat{\bm z})-\frac{3}{7}\mu^2 f \tilde{\mathcal{G}_2}\\
&\beta_1(k,\mu)\tilde 
\delta_1(\k,\hat{\bm z})
+\beta_2(k,\mu)
(\tilde{\delta}_1^2)^\perp (\k,\hat{\bm z}) \\
& +\beta_{\mathcal{G}_2}(k,\mu)
\tilde{\mathcal{G}_2}^{\perp}(\k,\hat{\bm z})
+\beta_3(k,\mu)
(\tilde{\delta}_1^3)^\perp (\k,\hat{\bm z})\,,
\end{split}
\ee 
where 
\be 
\mu = (\k\cdot \hat{\bm z})/k\,.
\ee 
The different bias operators 
are defined as follows.  
$\mathcal{G}_2$ is the tidal operator
with the unsymmetrized kernel
\be 
\label{eq:G2}
\begin{split}
\mathcal{G}_2(\k) & = \int_{\bm p} 
F_{\mathcal{G}_2}(\p,\k-\p)
\delta_1({\bm p})\delta_1(\k-{\bm p})\,,\\
 F_{\mathcal{G}_2}(\k_1,\k_2) & =\frac{({\bm k_1}\cdot \k_2)^2}{k_1^2 k_2^2}-1\,,
\end{split}
\ee 
$\int_{\k}\equiv \int \frac{d^3\k}{(2\pi)^3}$
and $\GG$ is the Galileon tidal operator, 
\be 
\begin{split}
F_{\GG}=\frac{4}{7}\left(1-\frac{(\k_1\cdot\k_2)^2}{k_1^2k_2^2}\right)
\left(\frac{((\k_1+\k_2)\cdot \k_3)^2}{(\k_1+\k_2)^2k_3^2}-1\right)\,.
\end{split}
\ee 
The above kernel should be used
in conjunction with the appropriate 
Fourier transform.
For a general cubic operator $\mathcal{O}^{(3)} $ we use the definition
\be 
\label{eq:gamma3def}
\mathcal{O}^{(3)} =  \int_{\k_1}\int_{\k_2}\int_{\k_3}\left(\prod_{i=1}^3\delta_1(\k_i)\right)(2\pi)^3\delta_D^{(3)}(\k-\k_{123})F_{\mathcal{O}^{(3)}}\,.
\ee 
The transfer functions are then 
measured from the
simulated density 
field snapshots $\delta^{\rm truth}$: 
\be 
\label{eq:trfreal}
\beta_i(k)=\frac{\langle \mathcal{O}^*{}^\perp_i(\k)  \delta_g^{\rm truth}(\k)\rangle'}{\langle
|\mathcal{O}^\perp_i (\k) |^2 
\rangle }'\,.
\ee 
The real space 
forward model is obtained
by setting $f=0$
in 
\eqref{eq:eft-field_rsd}
and 
removing the Zel'dovich field
$\delta_Z$ (whose contribution is 
 degenerate with $\b_1\tilde{\delta}_1$).
 The 
 transfer functions
 in this case
do not depend on the line-of-sight. 

Due to the presence of the 
stochastic components, 
the forward model always has noise,
whose power spectrum is given by
\be 
\label{eq:Perr}
P_{\rm err}(k,\mu) = 
\langle 
|\delta^{\rm EFT}(\k,\hat{\bm z}) - \delta^{\rm truth}(\k,\hat{\bm z})|^2
\rangle'
\ee 
In EFT, the theoretical prediction for the error power spectrum 
is given by 
\be 
\label{eq:rsd_stoch}
P_{\rm err}(k,\mu) = \frac{1}{\bar n}
\left(
1+\alpha_0+\alpha_1 \left(\frac{k}{k_{\rm S}}\right)^2
+\alpha_2\mu^2 \left(\frac{k}{k_{\rm S}}\right)^2
\right)\,,
 \ee 
where we have omitted higher derivative 
contributions. We choose $k_S=0.45~\hMpc$
following~\cite{Philcox:2021kcw}.

The bias parameters
and counterterms are extracted 
by fitting the transfer functions
and the error spectra with 
the EFT predictions using the methodology 
of ref.~\cite{Ivanov:2024xgb}
based on time-sliced perturbation
theory~\cite{Blas:2015qsi,Blas:2016sfa}.
We use $\kmax=0.4~\hMpc$
and $\kmax=0.2~\hMpc$
for real and redshift space
transfer function fits, 
respectively. We apply the 
same scale cuts to the 
noise power spectra. 
For Astrid the residual sample variance 
is very large at $\kmax=0.2~\hMpc$,
so we use $\kmax=0.4~\hMpc$ both in
real and redshift 
spaces.

\begin{figure*}[htb]
\centering
\includegraphics[width=0.99\textwidth]{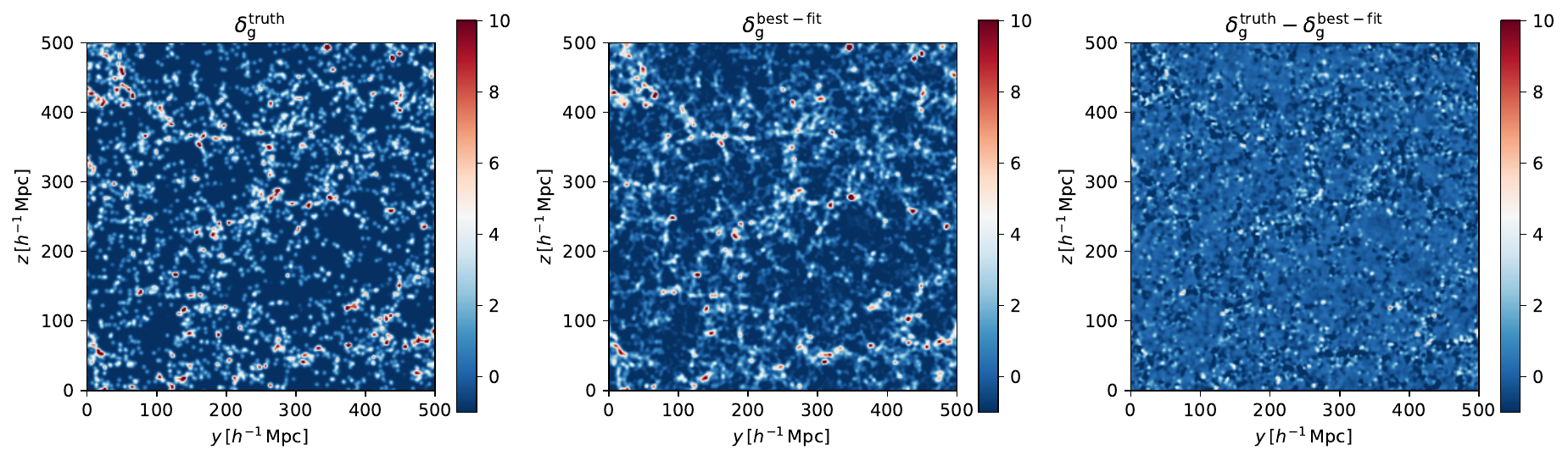}
\includegraphics[width=0.99\textwidth]{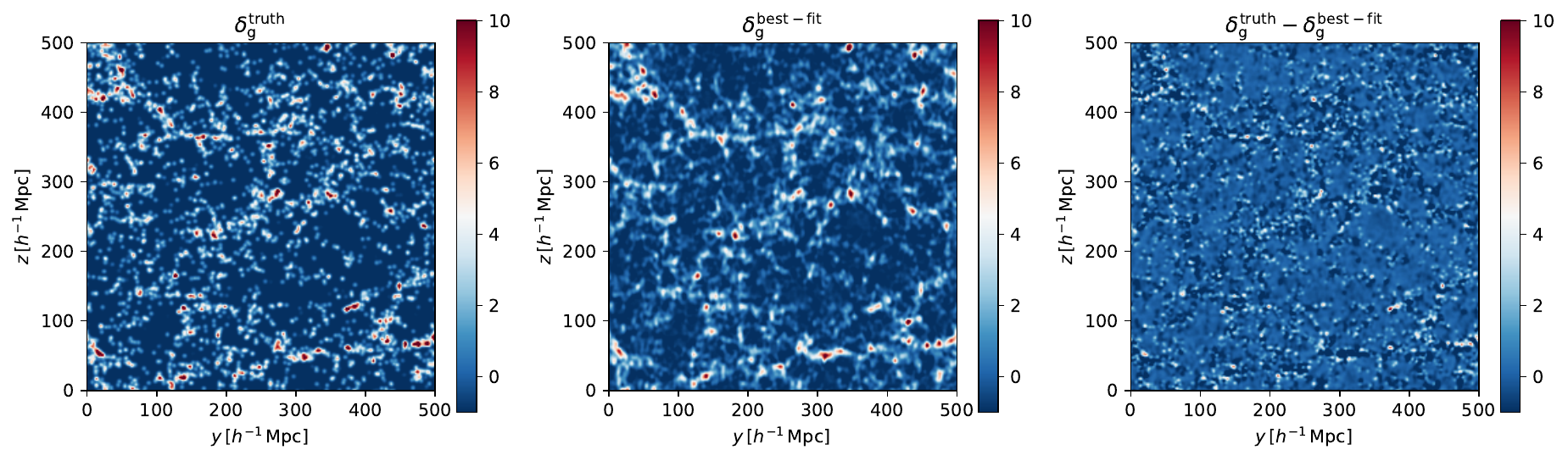}
   \caption{The MTNG DESI-like LRG density field in 
   real space (upper panel)
   and in redshift space (lower panel)
   from the MTNG simulation (left), its best field-level EFT fit (center), 
   and the residuals between the two (right). The square of the rightmost field gives the error power 
   spectrum. 
   For visualization purposes, 
   the field has been smoothed with a $R=2\,\Mpch$ 3D Gaussian filter. The depth of each panel is $\approx60\,\Mpch$. The redshift space distortions are applied along the $z$-axis.
    } \label{fig:field}
\end{figure*}
\begin{figure*}
\centering
\includegraphics[width=0.99\textwidth]{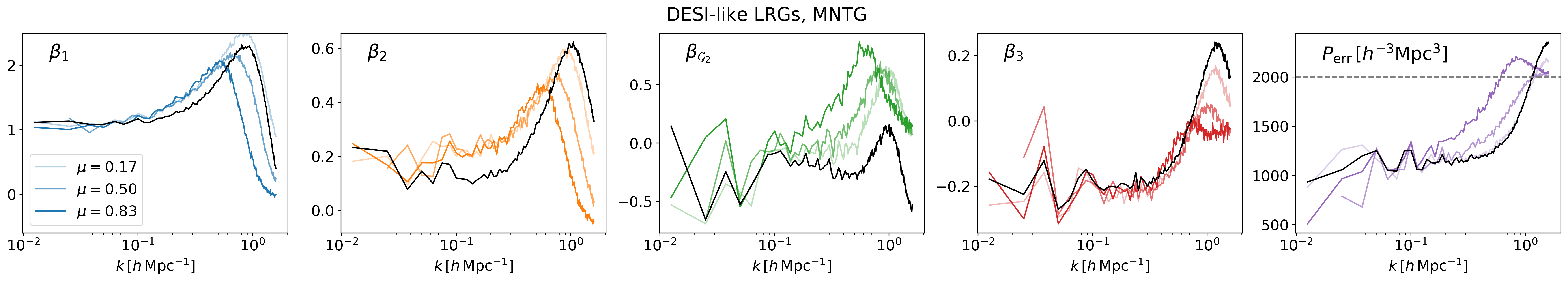}
\includegraphics[width=0.99\textwidth]{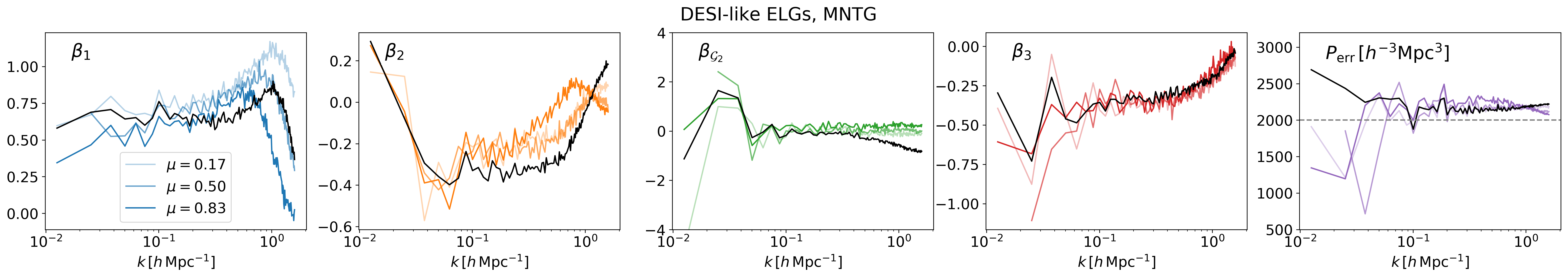}
   \caption{EFT model transfer functions 
   and noise power spectra
   for MTNG galaxies 
   in real space (black lines)
   and in redshift space, for three $\mu$ bins.
   Note that we subtract 1 and 0.5 from the real space transfer functions for $\beta_1$ and $\beta_{\G}$
   to match the low-$k$
   limit of the redshift space transfer functions. The dashed line in the rightmost panel
   depicts the $\bar n^{-1}$
    Poisson prediction.
    } \label{fig:transfer_mtng}
\end{figure*}

\begin{figure*}
\centering
\includegraphics[width=0.38\textwidth]{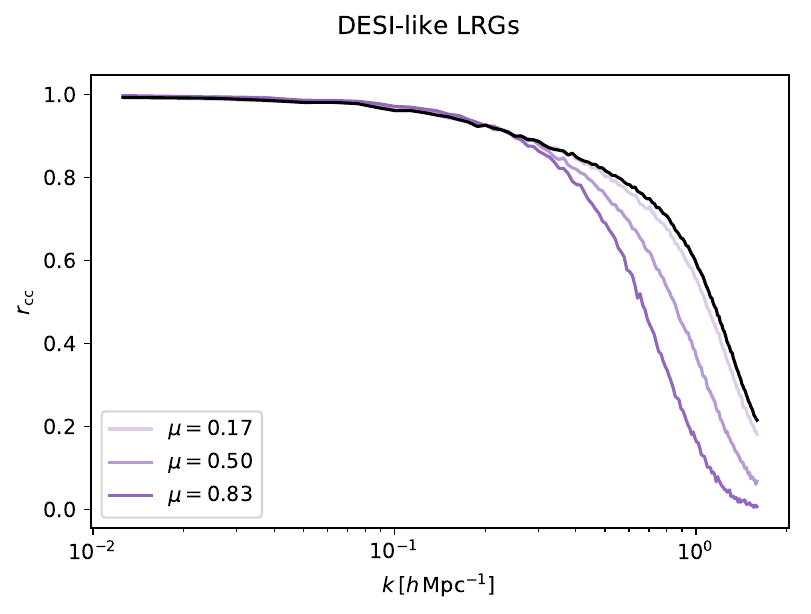}
\includegraphics[width=0.38\textwidth]{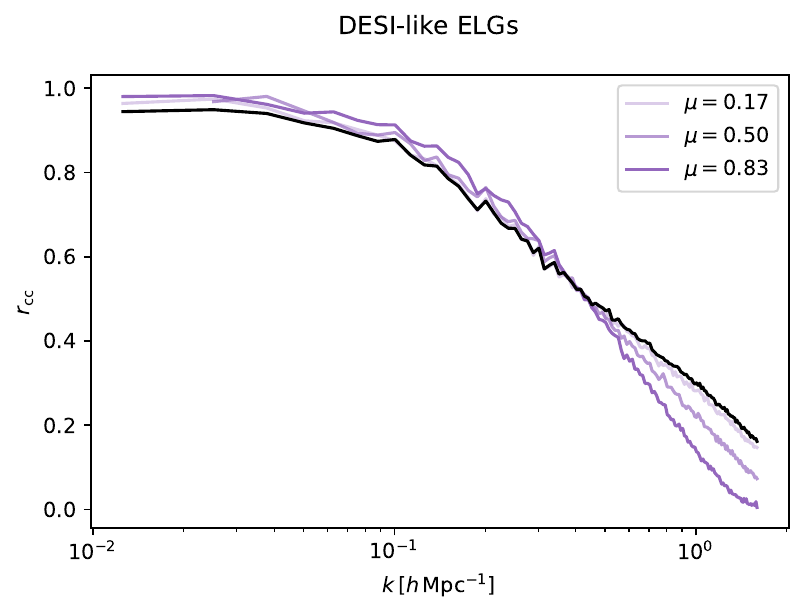}
   \caption{Cross-correlation coefficients $r_{cc}$ between the MTNG LRGs (left panel) 
   and ELGs (right panel).
   $r_{cc}$ measures
   the accuracy of the EFT model. For $k\lesssim 0.1~\hMpc$ it is closer to unity, implying that the galaxy distribution
   is almost entirely correlated 
   with the EFT operators built out of dark matter
   density and tidal fields. On large scales, the line-of-sight modes (non-zero $\mu$)
   increase the correlation
   with dark matter thanks to 
   the linear Kaiser effect. 
   The enhancement is proportional to $f/b_1$,
   which is larger for ELGs, owing to the higher redshift
   and smaller $b_1$.
    } \label{fig:rcc}
\end{figure*}


\section{Hydrodynamic galaxies in the field level EFT}
\label{sec:long_results}

Let us discuss now the 
results of our field-level 
measurements from
the hydrodynamic simulations.

\subsection{MTNG results}

The density fields 
and transfer functions extracted
from MTNG are shown in fig.~\ref{fig:field}
and fig.~\ref{fig:transfer_mtng}.
Visually, these 
transfer functions
are very similar to those
measured by us 
from HOD mocks~\cite{Ivanov:2024hgq}. 
An important parameter 
is the error power
spectrum $P_{\rm err}$
(see 
eq.~\eqref{eq:Perr})
as a function of $k$, 
which is related to the 
cross-correlation
coefficient, $r^2_{cc}=1-{P_{\rm err}}/{P_{\rm truth}}$, 
shown in fig.~\ref{fig:rcc} for DESI-like 
LRGs and ELGs. First, we see
that the error power spectrum
is approximately scale-independent,
in perfect agreement 
with the EFT predictions.
Second, from fig.~\ref{fig:rcc}
we see that 
the cross-correlation
coefficient is quite large
for wavenumbers $k\lesssim 0.4~\hMpc$, 
implying that EFT 
correctly reproduces 
the entire 
density field 
on quasi-linear scales.
The cross-correlation coefficient
is larger for the LRG sample because
they have a larger bias 
at the same number density
as ELGs. 
Third, we see that  
the scale-dependence
of $P_{\rm err}$
becomes sizeable only
around $k \simeq 0.5~\hMpc$,
just like 
in the HOD samples from~\cite{Ivanov:2024hgq}.
This suggests that 
the EFT model for the hydrodynamical
galaxies has 
a similar momentum reach 
$k_{\rm max}$
as the HOD-based
mocks. This is further confirmed 
by direct fits to the 
transfer functions,
which we use to extract 
the EFT parameters. 

\begin{figure*}
\centering
\includegraphics[width=0.99\textwidth]{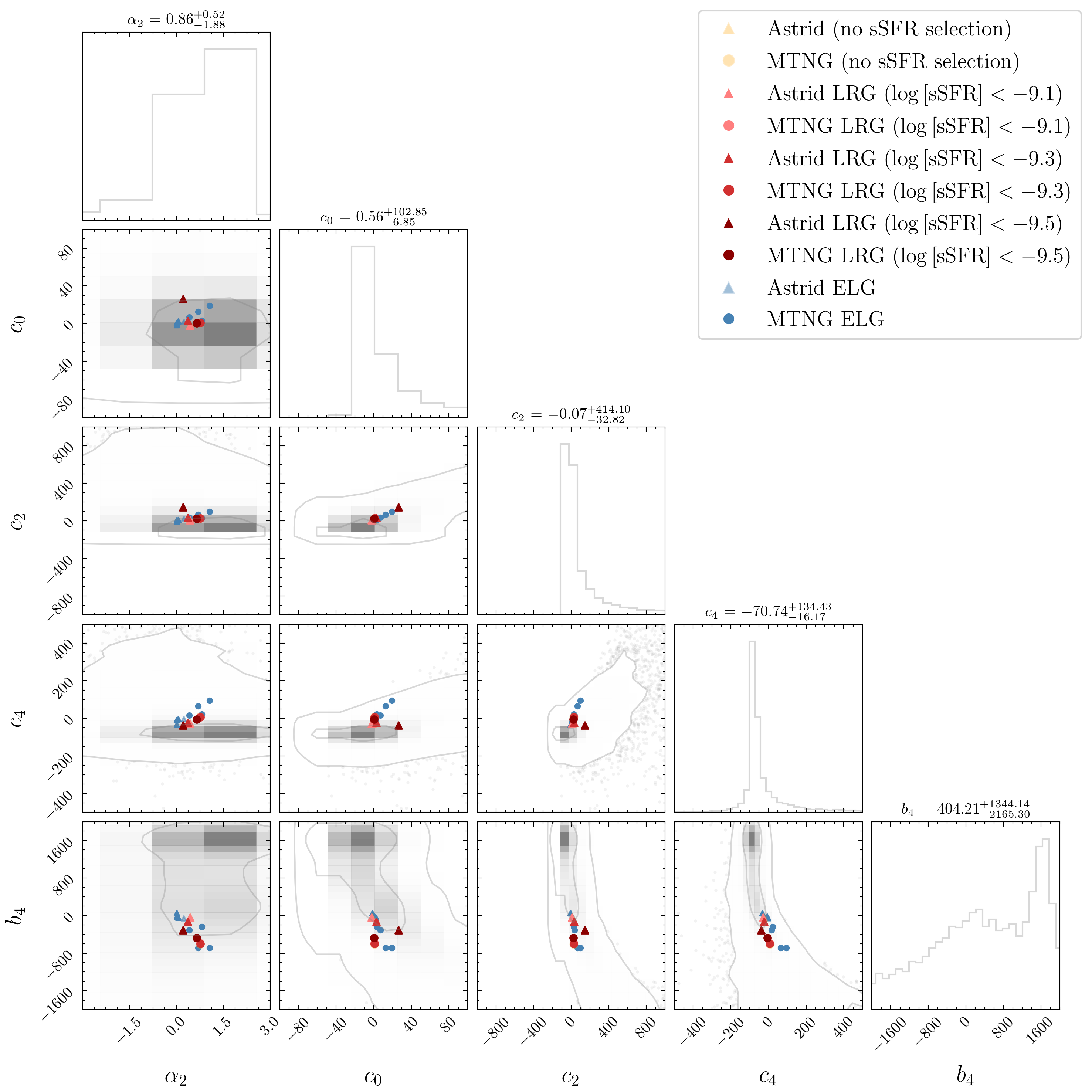}
   \caption{MTNG and Astrid redshift-space counterterms for LRGs and ELGs vs the LRG-HOD priors.
    } \label{fig:dist_rsd}
\end{figure*}

\begin{figure*}
\centering
\includegraphics[width=0.99\textwidth]{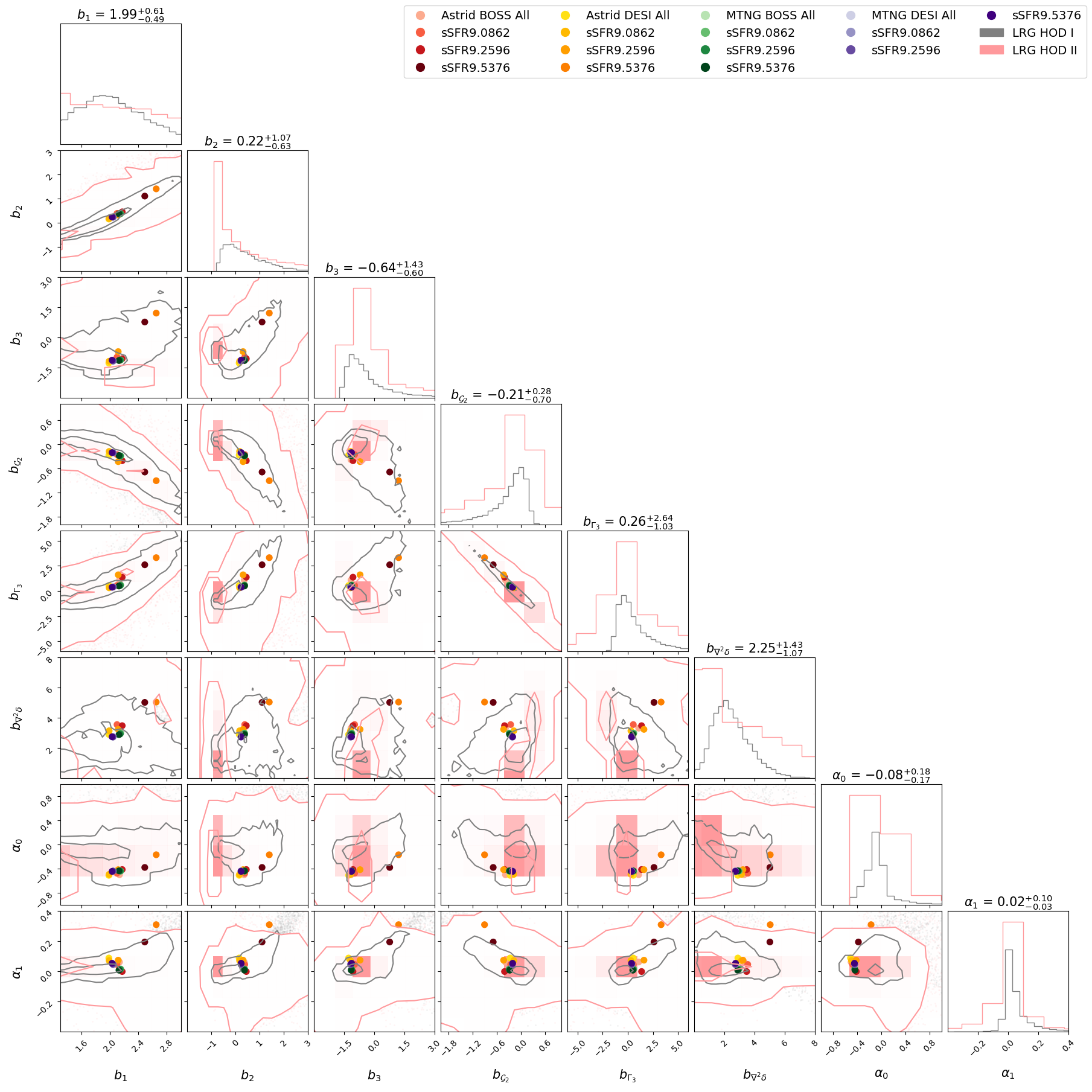}
   \caption{MTNG and Astrid real space EFT parameters of LRGs against two LRG-HOD-based distributions: LRG-HOD-I (gray) and LRG-HOD-II (pink).
  In contrast to LRG-HOD-I,
  LRG-HOD-II priors include
  concentration-dependent HODs 
  and have wider priors
  on the other HOD parameters. 
    } \label{fig:dist_bias_2}
\end{figure*}

For DESI-like galaxies, 
the real space EFT parameters, 
including galaxy 
bias parameters, are 
depicted by dots in fig.~\ref{fig:dist_bias}, 
while the redshift space 
counterterms are shown in 
fig.~\ref{fig:dist_rsd}.
In the background of this figure, 
we show measurements from 
the HOD-based mocks (HOD-LRG-I) from ref.~\cite{Ivanov:2024hgq}
for bias parameters
and ref.~\cite{Ivanov:2024xgb}
for redshift space counterterms. 
We display the measurements
for BOSS-like galaxies in fig.
\ref{fig:dist_bias_2}.
In addition, we show the priors
from the extended LRG-HOD-II sample.
Different selections in this figure
are referred to as sSFR$\tau_{\rm SFR}$,
e.g. sSFR9.0862 etc. LRGs/ELGs have 
larger/smaller $\tau_{\rm SFR}$
than the thresholds, respectively. 
We see that all the LRG data points
are fully consistent with the HOD-based 
priors from the LRG-HOD-II sample.

Let us discuss LRG datapoints first. 
We see that  
in general these are very 
consistent with the HOD-based
values. 
For most of the parameters,
the MTNG measurements 
lie within the 68\% 
probability 
density area.
The only parameter that is systematically 
outside of the $68\%$ confidence interval
is the sub Poisson stochasticity $\alpha_0$,
which still lies within the $95\%$ CL.
Let us note that
variations of the specific star
formation rate do not 
significantly change the values of LRG bias parameters.
This is consistent with the
previous 
studies of the $b_2$
and $b_1$
parameters
from 
IllustrisTNG~\cite{Barreira:2021ukk},
which considered a much
wider variation
of sSFR than us. 
The agreement between MTNG and TNG on the sSFR-dependence of $b_1$ and $b_2$
implies that the weak
response of 
MTNG LRGs to 
sSFR is a genuine
property of the TNG 
feedback. 
This point will be discussed further 
in comparison with Astrid galaxies. 
Comparing DESI and BOSS-like LRGs, we
see that the only relevant 
parameter 
that changes the values of EFT parameters in MTNG 
is the number 
density.

We emphasize that 
our comparison so far
has been based on 
the LRG-HOD-I sample of~\cite{Ivanov:2024hgq},
which does not include 
the concentration-dependent assembly bias parameters $A_{\rm cent},A_{\rm sat}$. Although these parameters  
can be included in HOD samples as in~\cite{Ivanov:2024xgb},
we see that they are
not necessary in order
to reproduce 
EFT parameters 
from MTNG.

Let us now move on to the redshift 
space counterterms, see fig.~\ref{fig:dist_rsd}. 
Overall, we see great consistency here is 
as well: all MTNG parameters 
lie within $95\%$ CL of the 
HOD-based parameter densities. 
Curiously, we notice 
a small 
spread of counterterm
values due to variations of 
sSFR, but 
it does not exceed 
the
residual 
statistical error
associated with
sample variance,
and hence it cannot be 
robustly 
interpreted 
as a genuine response
to sSFR. 
We also note that the MTNG LRGs 
exhibit somewhat weaker fingers-of-God
than the HOD galaxies,
and than the actual BOSS
samples~\cite{Ivanov:2019pdj,Ivanov:2021fbu}.

Finally, let us discuss 
MTNG ELGs, whose bias
parameters are shown
in both fig.~\ref{fig:dist_bias}
and fig.~\ref{fig:bias_elg}.
Looking at  fig.~\ref{fig:dist_bias},
we see that 
their bias parameters
exhibit a significant discrepancy with the HOD-LRG priors: 
the ELG data point lies outside of the 
HOD based density in 
the $b_2-b_1$, $b_1-b_2$ and $b_2-b_3$ 
planes. For all other real space parameters
the agreement is quite good. 
We have checked that considering 
a wider range of values for HOD parameters 
and including 
additional assembly bias 
parameters as in the LRG-HOD-II sample~\cite{Ivanov:2024xgb}
does not allow one to  
widen the density
of the local bias parameter towards
the ELG values.  
We will see shortly 
that the tension
is driven by the specific form
of the HOD for ELGs
which is not captured 
by the HOD model used in~\cite{Ivanov:2024hgq,Ivanov:2024xgb}
aimed at reproducing 
LRG clustering. This is confirmed in 
fig.~\ref{fig:bias_elg}, where 
we plot our MTNG ELG measurements
on top of the EFT parameters
following from the 
HMQ models.

As for the
redshift space ELG parameters, 
we see that the MTNG values are significantly smaller
than those of LRG-HODs. They 
are also typically smaller than 
the LRG redshift space counterterms 
from MTNG. This suggests
that ELGs may exhibit 
significantly 
weaker fingers-of-God than
LRGs in general. This is confirmed 
at the level of our HMQ-based
priors, see fig.~\ref{fig:rsd_elg}. There, we observe that the distribution 
of the conterterms 
from HMQ is much narrower
than that of the LRG-based HODs. 

Let us also note that there is an apparent $\sim 2\sigma$ tension
between the MTNG and HMQ-based
counterterm 
values in the $\alpha_2-b_4$
and $\alpha_2-c_4$ planes. While this tension
is not significant, it will be still interesting  
to understand its origin.

\begin{figure*}
\centering
\includegraphics[width=0.99\textwidth]{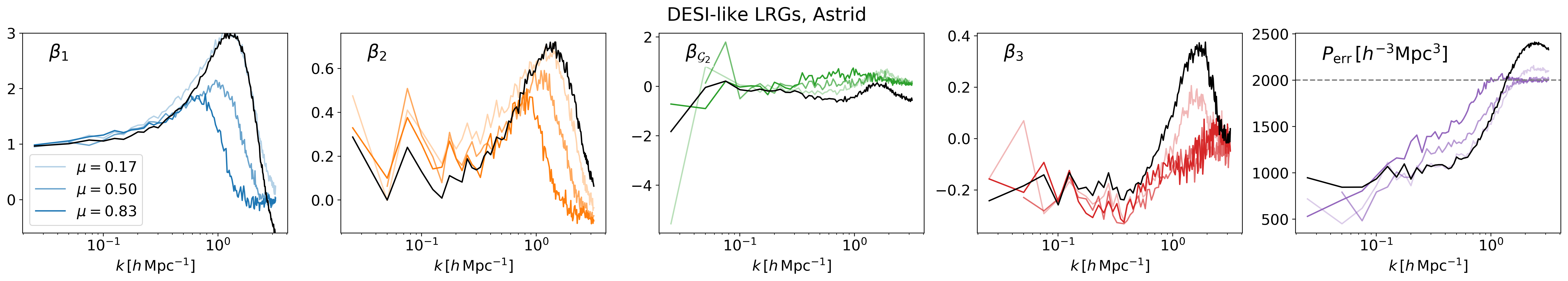}
\includegraphics[width=0.99\textwidth]{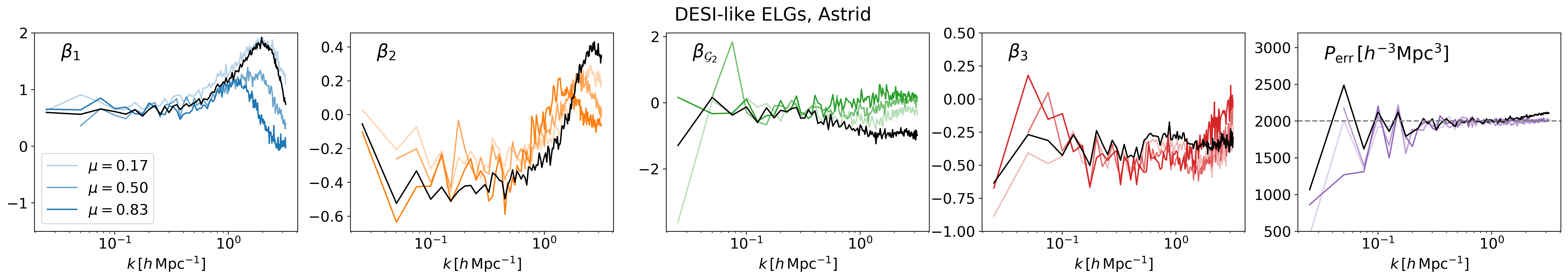}
   \caption{Same as fig.~\ref{fig:transfer_mtng}
   but for the Astrid simulation. 
    } \label{fig:astrid_transfer}
\end{figure*}

\subsection{Astrid results}

The Astrid EFT transfer functions
are shown in fig.~\ref{fig:astrid_transfer}.
As in the MTNG case, they look 
very similar to 
those of HOD-LRGs 
from ref.~\cite{Ivanov:2024xgb}.

Comparing the LRG datapoints
in fig.~\ref{fig:dist_bias}, one may notice 
that in contrast to MTNG, the sSFR 
cuts have 
a significant effect
on the Astrid galaxies. 
In addition, although most of the datapoints
still lie within $95\%$ CL, 
we see clear outliers for many EFT parameters. 
For instance, we see some noticeable 
outliers only in the $b_{\nabla^2\delta}$
and $\alpha_1$ planes.
For $b_{\nabla^2\delta}$, 
the deviations are quite natural
as the HOD mocks e.g. do not fully account
for the baryonic feedback,
which is captured in
hydrodynamic simulations.
We note however, that the 
tension w.r.t. the HOD values
of $b_{\nabla^2\delta}$
is not very significant. 
In particular, the $b_{\nabla^2\delta}$ distribution 
from HOD becomes noticeably 
wider once wider HOD-LRG-II priors
are considered,
as in fig.~\ref{fig:dist_bias_2},
in which case we find 
perfect agreement with Astrid. 

The different response 
of MTNG and Astrid galaxies
to sSFR cuts 
can be understood from
the arguments given in
Sec.~\ref{sec:gal_pops}
and from fig.~\ref{fig:selection}:
Astrid LRGs have a 
stronger dependence 
on sSFR than MTNG ones. 
While the distribution of LRGs
is approximately constant 
as a function of sSFR for low 
sSFR values for both Astrid and MTNG, 
the Astrid distribution changes dramatically as one lowers the sSFR (increases $\tau_{\rm sSFR}$).
The MTNG distribution, however, remains
approximately constant 
in the interval 
$\log(\text{sSFR/yr}^{-1})\in [-11,-9]$.
In other words, most of 
the LRG samples are 
concentrated far from the selection
threshold for low $\tau_{\rm sSFR}$,
but then their density changes
differently when lowering 
$\tau_{\rm sSFR}$, in agreement 
with expected differences in the feedback 
assumptions. 
This picture is consistent with 
the phenomenology seen in fig.~\ref{fig:dist_bias}: 
LRGs' biases for galaxies
with a low star 
formation rate 
from MTNG and Astrid
are quite similar,
and their response to sSFR  
is quite modest.
However, the discrepancy
between MTNG and Astrid 
grows with 
$\tau_{\rm sSFR}$.
Larger $\tau_{\rm sSFR}$
in Astrid corresponds to more massive,
more biased halos, which explains
why $b_{\nabla^2\delta}$ values
are large for these samples.

One may also notice some 
$\simeq 2\sigma$ outliers
in the $b_{\Gamma_3}$ and $\alpha_1$
planes, 
which suggest that spatial
non-localities are
quite significant for Astrid galaxies. 
The data points from 
Astrid exhibit a mild tension with HOD model 
considered in~\cite{Ivanov:2024hgq}, 
and their reproduction within these models may require fine-tuned 
HOD parameters. 
However, once we consider 
extended decorated 
LRG-HOD models from the 
LRG-HOD-II sample in fig.~\ref{fig:dist_bias_2},
all LRG data points 
become highly consistent 
with them.
Together with 
the $b_{\nabla^2\delta}$
behavior, 
this suggests
that an extended set of
HOD parameters as in~\cite{Ivanov:2024xgb} 
may be needed in
order to capture the higher derivative
bias of LRGs within the HOD framework.

Let us discuss now 
the Astrid ELGs. As in the MTNG case, 
their local bias parameters $b_1,b_2,b_3$
exhibit significant tensions
with the LRG-HOD model. 
Our results suggest that 
even in the context of the decorated
HOD models there is no way to 
reproduce the large-scale local bias 
parameters of ELG galaxies. 
The anomalous behavior, 
again, is captured by the HMQ-based
samples, see fig.~\ref{fig:bias_elg}.

As far as the ELGs' response
to sSFR is concerned, 
both Astrid and MTNG galaxies
have a significant response, 
consistent with 
their selection seen in fig.~\ref{fig:selection}.
Note however that the 
precise values of EFT parameters 
for the same sSFR cuts,
and the response as a function 
of sSFR
are slightly different between
MTNG and Astrid. 
Astrid's 
ELGs are less biased than MTNG ones
for the same sSFR cuts,
while the MTNG response 
to sSFR is smoother. This, again, 
can be understood 
from fig.~\ref{fig:selection}:
we see that 
the Astrid ELGs do not
response monotonically to 
lowering of sSFR as their distribution 
peaks around $\tau_{\rm sSFR}\approx 9.3$,
and hence the response to sSFR
slows down around this value. 
In contrast, the 
MTNG ELG distribution plateaus
only around $\tau_{\rm sSFR}\approx 10$,
growing towards this value monotonically. 
This explains why the 
Astrid results for $\tau_{\rm sSFR}=9.25$
and $\tau_{\rm sSFR}=9.54$
are almost the same,
while the MTNG 
EFT parameters keep changing
at the same pace as one 
raises $\tau_{\rm sSFR}$
in this region.

In addition, before reaching the plateau, the
density of Astrid ELGs is 
visibly larger than that of MTNG's (which can be appreciated
by comparing the black and gray contours
in fig.~\ref{fig:selection}).
This, again, can be traced
back to the fact that massive
galaxies in Astrid have 
a larger sSFR, implying 
larger abundance and 
consequently 
a lower linear bias of their ELGs,
consistent with our measurements. 

\begin{figure*}
\centering
\includegraphics[width=0.99\textwidth]{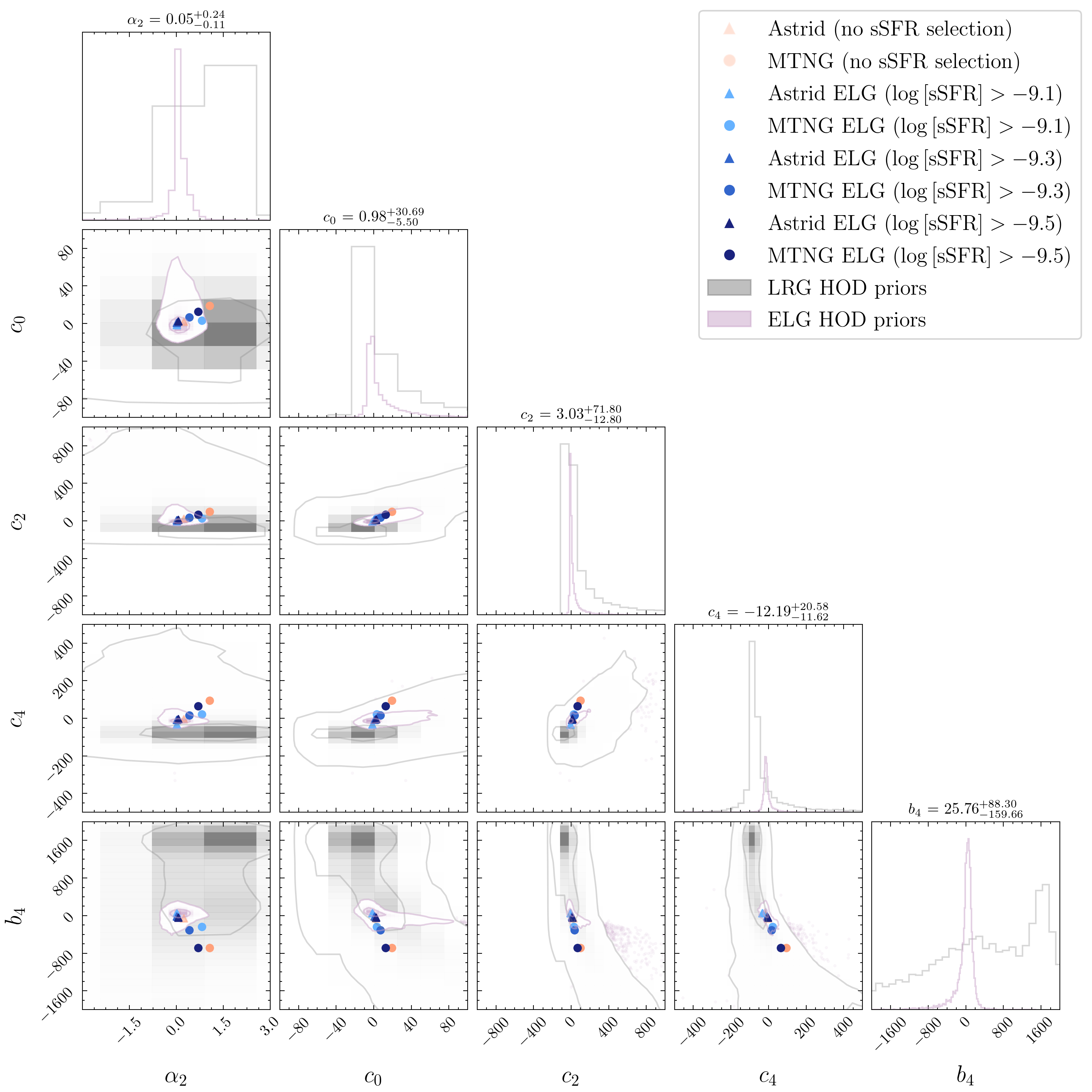}
   \caption{MTNG and Astrid redshift-space counterterms for ELGs vs the LRG-HOD and ELG-HMQ based distributions.
    } \label{fig:rsd_elg}
\end{figure*}

Finally, we note that the 
redshift-space EFT counterterms 
are very consistent with those of 
our HOD-LRG distribution in fig.~\ref{fig:dist_rsd}.
The comparison
against the HMQ priors
is given  
in fig.~\ref{fig:rsd_elg}.
As in the MTNG case, we see that the typical amplitude of the 
FOG ELG counterterms is smaller than those
of the HOD-LRG samples.
Note that the
$\log[\text{sSFR}]>-9.3$
samples
correspond to galaxies
with low star formation,
which do not actually
represent the observed ELGs. 
In terms of the selection
used in~\cite{Hadzhiyska:2022ugr},
they actually correspond to LRGs, 
and hence, strictly speaking, 
one should not compare them
with HMQ samples. 
In agreement with this argument,
we see that their counterterms 
are actually stronger than those
based on the HMQ samples. 
For the consistent choice of 
sSFR 
selection $\log[\text{sSFR}]>-9.1$, 
Astrid counterterms 
perfectly agree 
with the HMQ priors,
see fig.~\ref{fig:rsd_elg}.
In particular, 
RSD counterterms for Astrid ELGs 
match well the mean values
of the HMQ-based samples.

\subsection{Quadratic bias parameters of galaxies and halos}
\label{sec:bias}

\begin{figure*}
\centering
\includegraphics[width=0.49\textwidth]{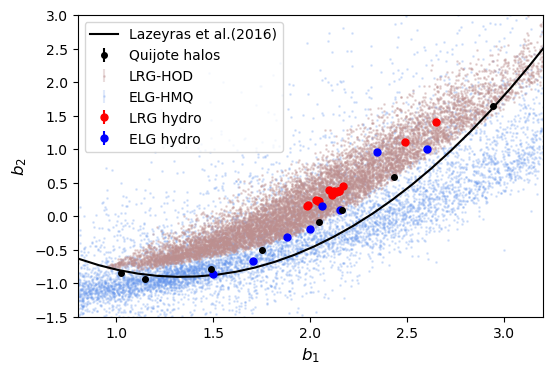}
\includegraphics[width=0.49\textwidth]{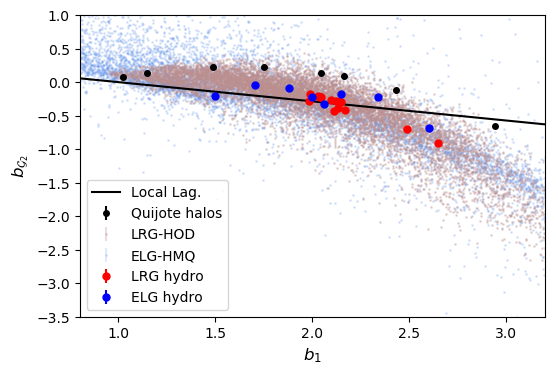}
   \caption{Quadratic bias parameters $b_2$ and $b_{\mathcal{G}_2}$
   as functions of $b_1$
   from the Quijote halos,
   LRG-HOD mocks, ELG-HQM mocks,
   and the hydro simulations MTNG and Astrid. The black curve
   represents the fit for $b_2(b_1)$ of halos
   from~\cite{Lazeyras:2015lgp} (left panel), 
   and the prediction of the 
   local Lagrangian model for $b_{\mathcal{G}_2}$ (right panel).
    } \label{fig:quad_bias}
\end{figure*}

One important result of our
work is the new measurement
of quadratic bias parameters
from the MTNG and Astrid simulations. 
The quadratic bias parameters
$b_2$ and $b_{\mathcal{G}_2}$
have been extensively studied in the literature, see e.g.~\cite{Baldauf:2012hs,Schmittfull:2014tca,Lazeyras:2015lgp,Lazeyras:2017hxw,Eggemeier:2021cam,Abidi:2018eyd,Barreira:2021ukk,Ivanov:2021kcd,Akitsu:2024lyt}. Let us compare our measurements 
of $b_2$ and $b_{\mathcal{G}_2}$
from hydro galaxies with
this literature 
and with the values extracted
from our halo occupation
models. 

Our main results 
are summarized in Fig.~\ref{fig:quad_bias}.
For simplicity, we combine
the Astrid and MTNG 
measurements. 
First, let us discuss 
the local bias parameter $b_2$
as a function
of the linear bias parameter $b_1$. 
The first point of comparison 
is the $b_2(b_1)$ function of 
dark matter halos. It is shown
in the left panel of Fig.~\ref{fig:quad_bias}
as a numerical fit from the analysis of~\cite{Lazeyras:2015lgp},
and also as direct measurements
of $b_2(b_1)$ values from the 
\texttt{Quijote}
simulation~\cite{Villaescusa-Navarro:2019bje} from~\cite{Ivanov:2024xgb}.
We see that the LRG $b_2(b_1)$ 
curve lies systematically above
the halo curve for $b\sim 2.2$, 
but the difference between them
decreases for larger values of $b_1$.
This trend was first pointed 
out by~\cite{Eggemeier:2021cam}. 
The hydro ELG, however, 
agree well with the pure 
halo predictions. 
The second point of comparison
is between the hydro galaxies 
and the HOD/HMQ-based samples.
There we see a high level 
of consistency: both LRG 
and ELG
from our hydro simulations 
match well the 
halo-based
values. 

The intricate behavior of 
galactic $b_2$ values 
can be easily understood as a 
consequence of the specific HOD shape
for different galaxies. 
In general, the $b^g_2$
of galaxies is given by the 
$b^h_2(M)$ of halos, weighted 
with the halo mass function
and the HOD~\cite{Desjacques:2016bnm,Akitsu:2024lyt}:
\be 
\label{eq:biasHOD}
b_2^g = \frac{1}{\bar n_g}
\int dM \frac{d\bar n_h}{dM}
\langle N\rangle (M)b^h_2(M)\,,
\ee 
where 
$\langle N \rangle= \langle N_c \rangle+\langle N_s \rangle$
is the average HOD,
$\bar n_g$ is the galaxy number
density,
and $\frac{d\bar n_h}{dM}$
is the halo mass function. 
Let us discuss the LRG case first.
Let us ignore the satellites, 
assume for simplicity 
that 
the 
HOD can be approximated 
as a Heaviside step function,
$\langle N_c\rangle (M)\approx \Theta_H(M-M_{\rm cut})$,
and approximate $b_2^h(M)$
around its minimum
as a parabola, 
\[
b_2^h(M)\approx -b_2^{\rm min}+b'_2(\ln M-\ln M_{0})^2\,,
\]
with $b_2^{\rm min},b'_2>0$.
For the Press-Schechter 
halo mass function in a Planck-like cosmology 
$M_0\approx 10^{12.4} h^{-1}M_{\odot}$
at $z=0.5$
($M_0(z=1) \approx 10^{12}h^{-1}M_{\odot}$)~\cite{Tinker:2010my,Desjacques:2016bnm}.
This implies
\be 
\label{eq:b2appr}
b_2^g \approx -b_2^{\rm min}+
b'_2
\frac{1}{\bar n_g}
\int_{M_{\rm cut}} dM 
\frac{d\bar n_h}{dM}
(\ln M-\ln M_{0})^2\,.
\ee 
Obviously, the rightmost term above is greater than zero, 
so that $b_2^g$ will always be 
greater than 
the minimal value of $b_2$
of halos
$-b_2^{\rm min}$.
Let us further specify now 
that $\frac{d\bar n_h}{dM}\propto M^{-n}$ for $M<M_*$, which is a good approximation for light halos in a $\Lambda$CDM universe 
within the Press-Schechter model
(for which $n\approx 2$ and $M_*\sim 10^{14}M_\odot/h$). 
Then the integral in eq.~\eqref{eq:b2appr}
is dominated by its lower limit, 
which gives: 
\be 
\begin{split}
& b_2^g \approx -b_2^{\rm min}+
b_2'{\ln^2x} +
\frac{2b'_2}{(n-1)^2}\left[1+(n-1)\ln x\right]\,,
\end{split}
\ee 
with
$x\equiv {M_{\rm cut}}/{M_0}$.
Comparing this with
the relevant halo value, 
\be 
b_2^h(M_{\rm cut})=-b_2^{\rm min}+b'_2{\ln^2 x}\,,
\ee 
and recalling that $M_{\rm cut} \gtrsim M_0$ for our baseline LRG-HOD-I sample,
we see that this formula 
correctly reproduces the additional enhancement of galactic $b_2$:
\be 
\label{eq:deltab}
\begin{split}
b_2^g - b_2^h &= b'_2\frac{2}{(n-1)^2}\left[1+(n-1)\ln x\right]\\
&\approx  2b'_2(1+\ln(M_{\rm cut}/M_0))~\,,
\end{split}
\ee 
In the case $M_0\gtrsim M_{\rm cut}$ relevant for some mocks in our extended HOD-LRG-II sample, we 
still 
have an enhancement,
albeit somewhat smaller, because the logarithm in the r.h.s.~above is negative but bounded from below, $\ln(M_0/M_{\rm cut})> -1$.

\begin{figure*}
\centering
\includegraphics[width=0.99\textwidth]{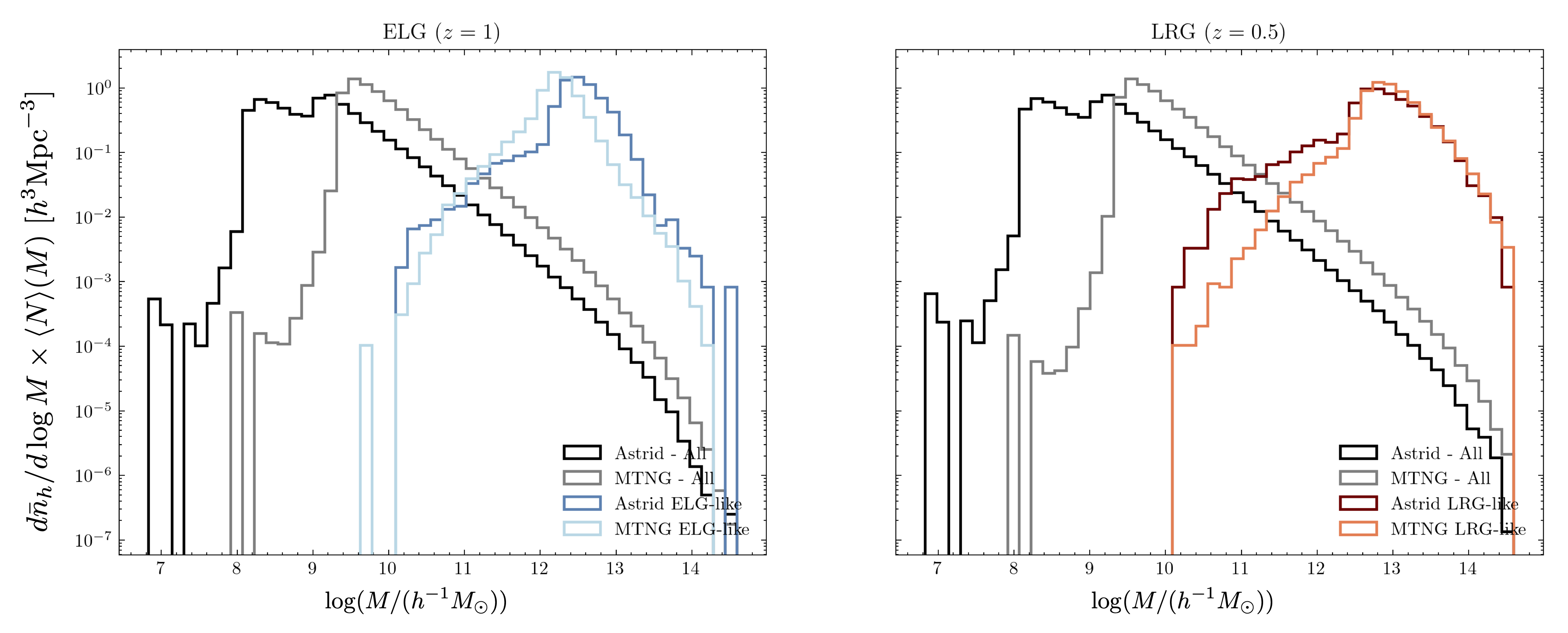}
   \caption{The mass 
   function of halos hosting
   the hydrodynamic galaxies, i.e. $\frac{d\bar n_h}{d\log M}\langle N \rangle(M)$
   where $\langle N \rangle(M)$ is the halo occupation distribution.
   The selection of galaxies matches the one 
   displayed in fig.~\ref{fig:selection}.
    } \label{fig:hmf}
\end{figure*}

Note that our derivation
above is very simplistic, yet it shows that key to understanding the LRGs' $b_2$ is the shape of their HOD, plus the fact that 
these galaxies 
typically reside
in halos whose masses are 
larger than $M_0$ that minimizes $b^h_2(M)$. 
We have ignored the satellites in our discussion, but their growing with mass HOD shape
will always additionally 
enhance
the total $b_2$ of the sample.

The total effective weight of the galaxy
bias 
$\frac{d\bar n_h}{dM}\langle N \rangle$
that appears
in the EFT parameter relations
such as eq.~\eqref{eq:biasHOD}
is quite broad
for $M>M_0$, see Fig.~\ref{fig:hmf}
that displays the mass function
of halos hosting the Astrid and 
MTNG galaxies. 
In Fig.~\ref{fig:hmf} the weight peaks at  $M\sim 10^{13}h^{-1}M_\odot$
and then extends 
towards larger masses.
This implies
that the HOD weighting 
always produces a positive 
correction to $b_2$, 
which 
explains the observed enhancement.

As far as the ELG are concerned, 
their HOD weighted by the halo mass function has a narrow peak (around $M\sim 10^{12}h^{-1}M_\odot$ in fig.~\ref{fig:hmf}), 
which can be crudely 
approximated 
as
a Dirac delta-function, 
$\langle N \rangle \frac{d\bar n_h}{dM}\propto\delta_D(M-M_{\rm cut})$.
In this case we obviously reproduce the desired leading order 
result
$b_2^g(M_{\rm cut)}\approx b_2^h(M_{\rm cut)}$. 
Going beyond the delta-function limit, 
we can see that the size of the correction
to $b_2$ is determined by the spread
of the HOD bias weighting function
$\frac{d\bar n_h}{dM}\langle N \rangle$.
This, in addition 
with the ELGs populating 
lighter halos, 
results in the spread   
of the $b_2$ values around the halo
curve for ELG which can be observed in  
Fig.~\ref{fig:quad_bias}.
Thanks to 
the ELG HOD weight,
the fluctuations 
around the halo values
can be both positive 
and negative,
as 
can be seen in fig.~\ref{fig:quad_bias}.

Let us now discuss the tidal bias $b_{\mathcal{G}_2}$
as a function of $b_1$, 
shown in the right panel of 
Fig.~\ref{fig:quad_bias}.
The first point of comparison
is the dark matter halo values
from \texttt{Quijote}. 
We see that in general the 
galaxy $b_{\mathcal{G}_2}$ values
lie below the 
halo values. 
The second point of comparison
is with the local Lagrangian
prediction $b_{\mathcal{G}_2}^{\rm Local~Lag.}=-(2/7)(b_1-1)$. There, we see that $b_{\mathcal{G}_2}$
of our hydro galaxies
indeed matches well the local Lagrangian
model for $b_1\lesssim 2.5$. This can be 
contrasted with the 
halos, 
which appear in tension 
with the local Lagrangian model
for $b_1\sim 2$,
in agreement with~\cite{Abidi:2018eyd,Eggemeier:2021cam}.
Finally, 
we observe a high level of consistency
between the hydro galaxies
and the HOD/MHQ models.

\subsection{Optimal HOD from EFT
parameters}

\begin{figure*}
    \centering
    \includegraphics[width=0.98\linewidth]{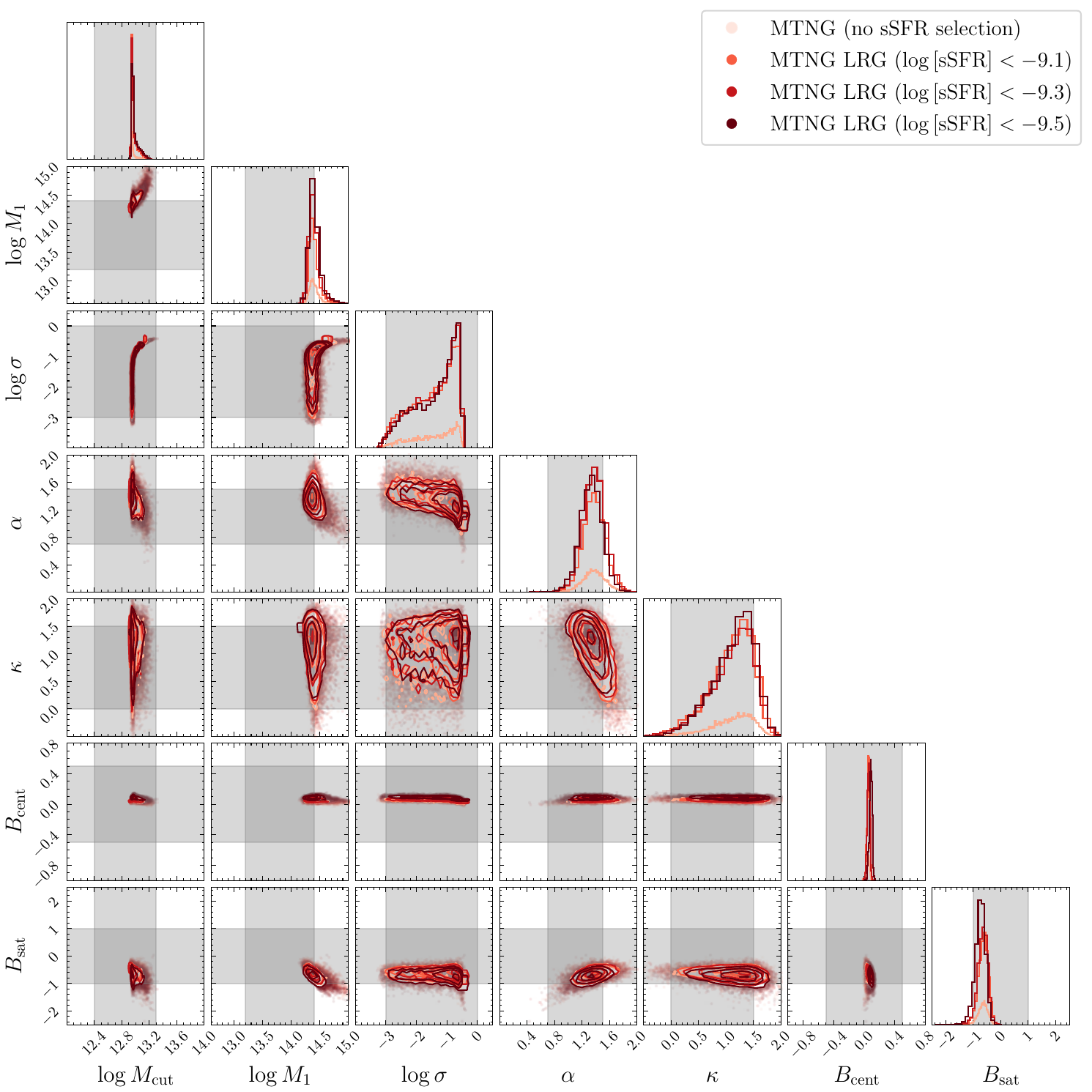}
    \caption{Inferred distribution of LRG-HOD parameters for MTNG as translated with our learned conditional distribution and using measured EFT parameters. The gray bands represent the range of priors used on HOD parameters in the NF's training set. See refs.~\cite{Ivanov:2024hgq,Ivanov:2024xgb} for further details on the underlying data sets.}
    \label{fig:mtng-eft-to-hod}
\end{figure*}

\begin{figure*}
    \centering
    \includegraphics[width=0.98\linewidth]{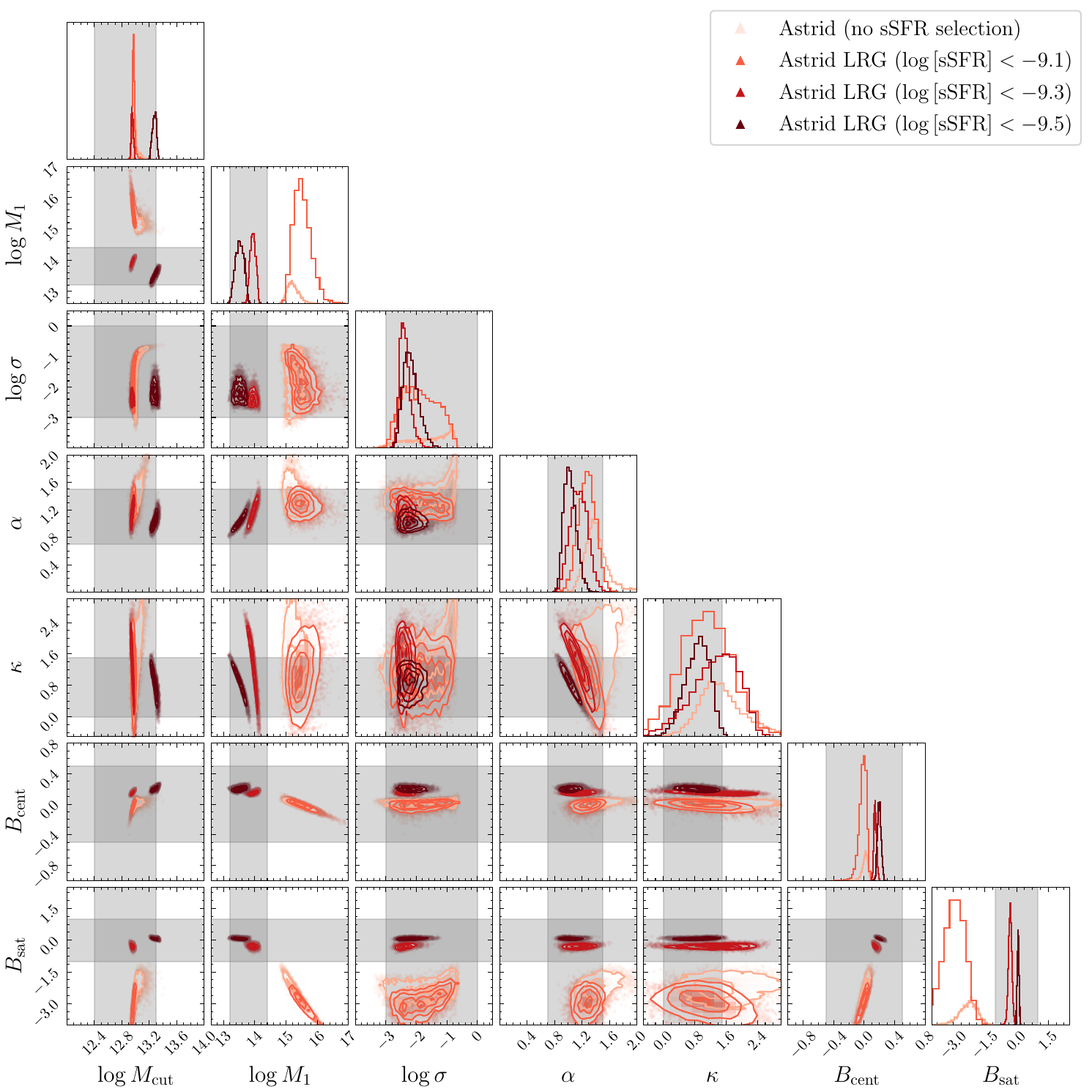}
    \caption{Same as fig.~\ref{fig:mtng-eft-to-hod} but for the Astrid simulation.}
    \label{fig:astrid-eft-to-hod}
\end{figure*}

\begin{figure*}
    \centering
    \includegraphics[width=0.98\linewidth]{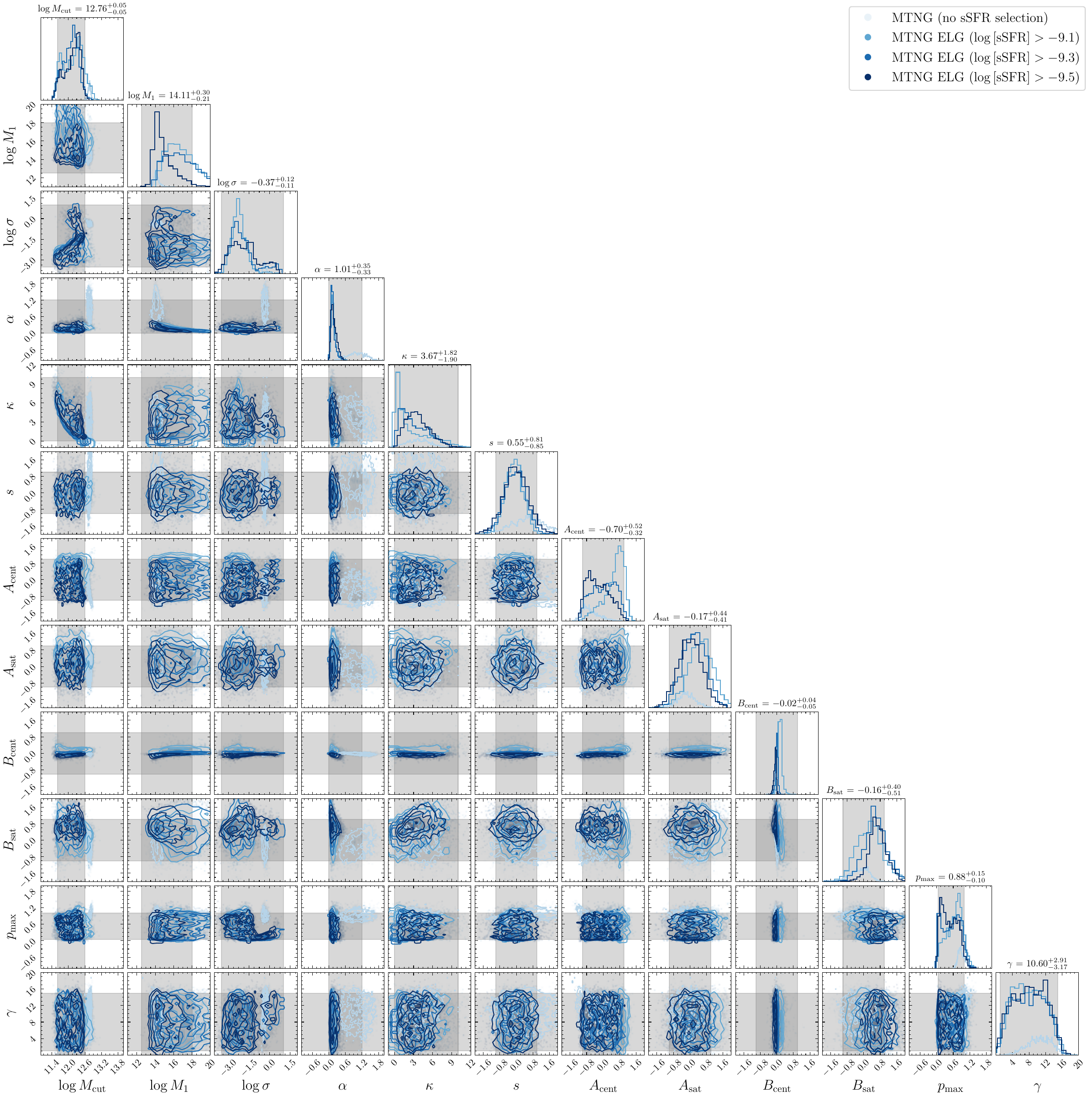}
    \caption{Inferred distribution of HOD parameters for MTNG ELGs as translated with our learned conditional distribution and using measured EFT parameters. The gray bands represent the range of priors used on HOD parameters in the NF's training set (Eq.~\ref{eq:elg-prior}).}
    \label{fig:mtng-elg-eft-to-hod}
\end{figure*}

\begin{figure*}
    \centering
    \includegraphics[width=0.98\linewidth]{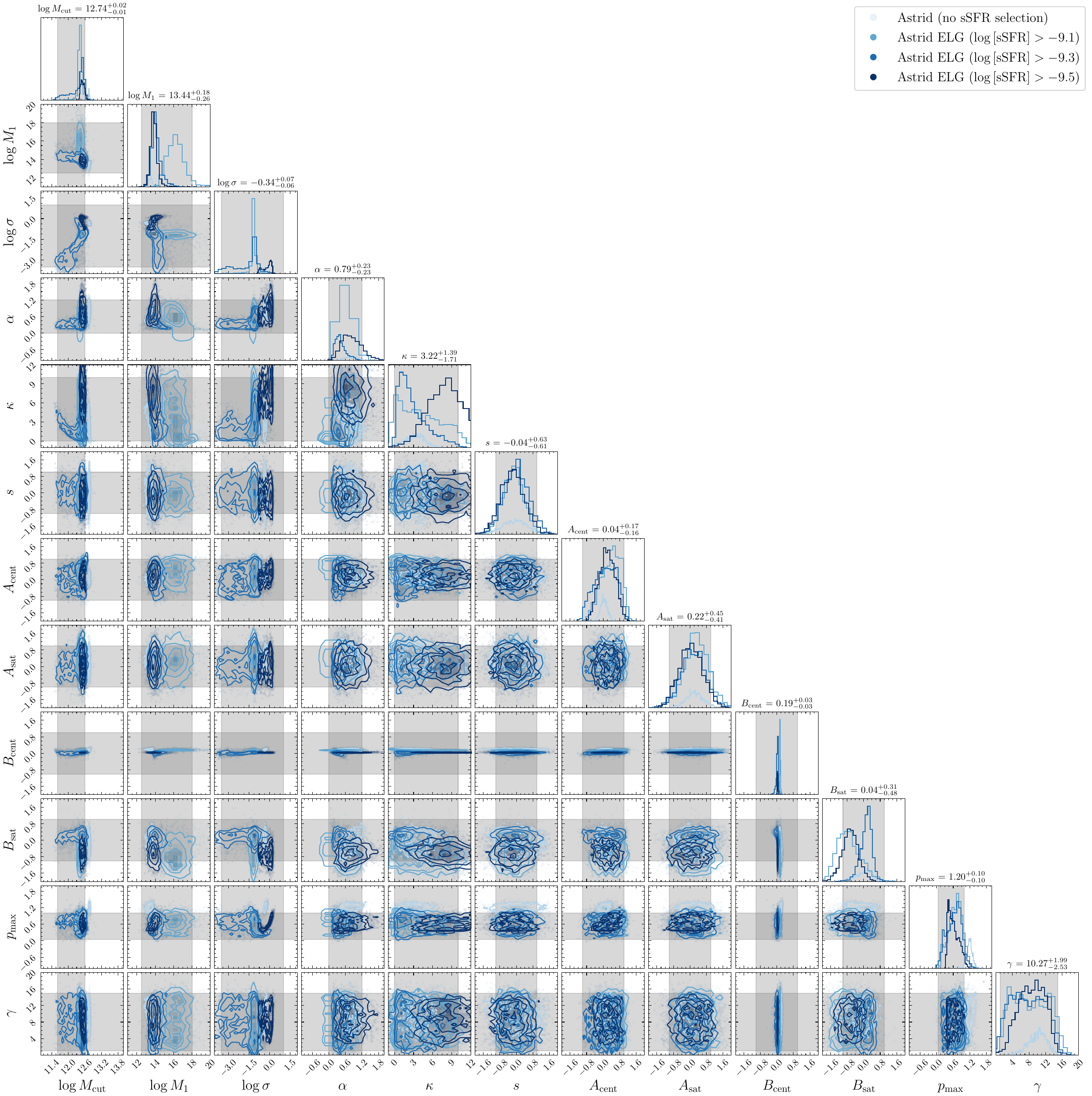}
    \caption{Same as fig.~\ref{fig:mtng-elg-eft-to-hod} but for the Astrid simulation.}
    \label{fig:astrid-elg-eft-to-hod}
\end{figure*}

We have found excellent agreement between the EFT parameters of hydro galaxies derived from hydrodynamic simulations and those predicted by HOD models. Using the paired distribution of EFT and HOD parameters provided in refs.~\cite{Ivanov:2024hgq,Ivanov:2024xgb}, one can construct a conditional distribution to translate measured EFT parameters from hydrodynamic simulations to the corresponding HOD parameters. This approach offers a computationally inexpensive alternative to directly matching the measured HOD shape in hydrodynamic simulations, enabling a quick estimation of the optimal HOD parameters required to fit hydrodynamic data. To illustrate the utility of this method, we present an example of inferring the HOD parameters needed to reproduce the EFT parameters of hydrodynamical simulations.

To learn the conditional distribution $p(\theta_{\rm HOD}|\theta_{\rm EFT})$, we trained a normalizing flow (NF) model using 10,500 mock HOD catalogs generated with \texttt{AbacusSummit} N-body simulations, each associated with measured EFT parameters. 
As a first step, we focus 
on the paired samples of a HOD-LRG-I model and real space EFT parameters from~\cite{Ivanov:2024hgq}.
The NF model was trained for 40,000 steps, during which validation and training losses were carefully monitored to ensure the absence of overfitting. Upon completion, the NF model effectively captured the intricate relationships between the HOD and EFT parameter spaces, enabling efficient and accurate mapping between them.

Building on the strong agreement between the measured EFT parameters of hydro galaxies and HOD models demonstrated earlier, no significant domain shift is expected when using hydro galaxy EFT parameters as input. To test this, we sampled the conditional distribution 5,000 times for each input, producing inferred distributions of HOD parameters for MTNG and Astrid. The results are shown in figs.~\ref{fig:mtng-eft-to-hod} and \ref{fig:astrid-eft-to-hod}, respectively. Broadly, the generated conditional distributions align well with the HOD priors. However, for Astrid, two samples extend noticeably beyond the priors, likely due to an unresolved degeneracy between
the satellite HOD 
parameters $\log{M_1}$ and $B_{\rm sat}$, 
clearly visible in fig.~\ref{fig:astrid-eft-to-hod}.
This example demonstrates the practical utility of the trained normalizing flow model, providing a computationally efficient and robust method for estimating HOD parameters while addressing the challenges of traditional matching techniques.

Note that both for Astrid and MTNG
our predictions 
for the halo mass threshold of centrals $\log M_{\rm cut}$, and their 
density-dependent assembly bias
parameter $B_{\rm cent}$
are much narrower than the priors. 
$B_{\rm cent}$ is 
consistent with zero
for all hydro samples.
The values of $M_{\rm cut}$
agree well with the halo mass
function corresponding to the 
host halos, see fig.~\ref{fig:hmf}.

Beyond LRGs, we have also repeated this analysis using paired samples of the HMQ-ELG model and real-space EFT parameters. In particular, we employed the ELG model from eq.~\ref{eq:elg-prior}, excluding $\alpha_s$ and $\alpha_c$ from the training sample. This approach realizes a mapping from 9 EFT parameters to 12 HOD parameters. After training our NF architecture for 40,000 steps with 10,500 mocks, we generated the inferred distribution of HOD parameters for MTNG and Astrid by conditioning on their measured EFT parameters. These results are shown in figs.~\ref{fig:mtng-elg-eft-to-hod} and \ref{fig:astrid-elg-eft-to-hod}, respectively. We find that our results are broadly consistent with the range of priors on HOD parameters used during training. However, it is worth noting that samples without sSFR selection appear somewhat discordant with the prior. This is expected, as such a choice of cut primarily represents LRGs. This is also reflected in the larger values of $M_\mathrm{cut}$ observed.

\section{Discussion}
\label{sec:disc}

In this work we have studied
the accuracy of modeling 
hydrodynamic galaxies 
at the field level 
in the context of the EFT forward model. 
Our main results are summarized in Section~\ref{sec:summ}. Let us now discuss
the main extensions of our analysis. 

From the theory point of view, it will be 
interesting to use the semi-analytic 
galaxy formation approaches,
such as~\cite{Marinucci:2023jag},
to compute the EFT 
parameters of red and blue galaxies.

It will be important 
to study the EFT parameters
of other types
of tracers, 
such as 
e.g. 
the Lyman alpha forest~\cite{Ivanov:2023yla,Ivanov:2024jtl}, quasars~\cite{Chudaykin:2022nru},
or intrinsic alignments~\cite{Chen:2024vvk}.
The latter is particularly 
important as the hydrodynamic 
simulation is currently the only 
robust approach to producing 
simulation-based
priors for intrinsic alignments of 
galaxies. 
This work is currently underway. 

Our work confirms the validity the 
field-level HOD-based
priors developed 
and applied to the BOSS data in~\cite{Ivanov:2024xgb}.
This work has found that the application
of HOD-based priors results in 
a $\sim 5\sigma$ tension
with the \textit{Planck} $\Lambda$CDM
model~\cite{Planck}. 
Our current study thus reinforces
this tension in the BOSS data, 
which was also
independently 
confirmed with BOSS-CMB lensing
cross-correlation data in~\cite{Chen:2022jzq,Chen:2024vuf}. 
It will be interesting to understand
the role of EFT/HOD parameters 
in this tension.

Finally, it will be important to 
extend our analysis to 
other hydrodynamical 
simulations 
and other variants
of the subgrid models. 
The ultimate goal
of this research direction
is to generate priors
for EFT-parameters
based on different hydrodynamic models. 
This leave
this and other research directions
for future exploration.

\section*{Acknowledgments}
MI thanks Matias Zaldarriaga 
for valuable  discussions
and comments on the draft.
This work is supported by the National Science Foundation under Cooperative Agreement PHY-2019786 (The NSF AI Institute for Artificial Intelligence and Fundamental Interactions, \url{http://iaifi.org/}). This material is based upon work supported by the U.S. Department of Energy, Office of Science, Office of High Energy Physics of U.S. Department of Energy under grant Contract Number  DE-SC0012567. AO acknowledges financial support from the Swiss National Science Foundation (grant no CRSII5{\_}193826). MWT  acknowledges financial support from the Simons Foundation (Grant Number 929255).
YN is supported by ITC postdoctoral fellowship. CH-A acknowledges support from the Excellence Cluster ORIGINS funded by
the Deutsche Forschungsgemeinschaft (DFG, German Research Foundation)
under Germany's Excellence Strategy -- EXC-2094 -- 390783311.
RK acknowledges support of the Natural Sciences and Engineering Research Council of Canada (NSERC) through a Discovery Grant and a Discovery Launch Supplement (funding reference numbers RGPIN-2024-06222 and DGECR-2024-00144) and York University's Global Research Excellence Initiative.  LH acknowledges support by the Simons Collaboration on ``Learning the Universe''. SB acknowledges funding from a UK Research \& Innovation (UKRI) Future Leaders Fellowship [grant number MR/V023381/1].



\bibliography{short.bib}

\end{document}